\documentclass[%
 reprint,
 superscriptaddress,
 amsmath,amssymb,
 aps,
 prd,
]{revtex4-2}

\usepackage[breaklinks=true]{hyperref} 
\usepackage{dcolumn}
\usepackage{bm}
\usepackage{hyperref}
\usepackage{amsmath}	
\usepackage{amssymb}	
\usepackage{mathtools} 
\usepackage[thinc]{esdiff}
\usepackage{cleveref}
\usepackage{natbib}
\usepackage{color}
\usepackage{diagbox}
\usepackage{siunitx}
\makeatletter
\renewcommand\p@paragraph{\thesection.} 
\makeatother

\usepackage{graphicx}
\graphicspath{ {./figures/} }

\begin{document}

\let\vec\boldsymbol
\def\E{\mathrm{E}}
\def\inc{I}
\def\mstar{M_{\bullet}}
\def\real{\operatorname{Re}}
\def\imag{\operatorname{Im}}
\def\L{\mathrm{L}}


\newcommand{\araa}{Annual Review of Astronomy and Astrophysics}
\newcommand{\apjs}{Astrophysical Journal, Supplement}
\newcommand{\apjl}{Astrophysical Journal, Letters}
\newcommand{\aap}{Astronomy and Astrophysics}
\newcommand{\icarus}{Icarus}
\newcommand{\mnras}{Monthly Notices of the RAS}
\newcommand{\physrep}{Physics Reports}
\newcommand{\na}{New Astronomy}
\newcommand{\aj}{Astronomical Journal}
\newcommand{\pasp}{Publications of the ASP}
\newcommand{\actaa}{Acta Astronomica}
\newcommand{\pasj}{Publications of the Astronomical Society of Japan}
\newcommand{\zap}{Zeitschrift fuer Astrophysik}

\title{Numerical approach to second-order canonical perturbation theory \\ 
        in the planetary 3-body problem: Application to exoplanets }


\author{Aya Alnajjarine}
\thanks{Corresponding author. \href{mailto:aya.alnajjarine@obspm.fr}{aya.alnajjarine@obspm.fr}}
\affiliation{IMCCE, CNRS, Observatoire de Paris, Universit\'{e} PSL, Sorbonne Universit\'{e} \\ 
77 Avenue Denfert-Rochereau, 75014 Paris, France}
\author{Federico Mogavero}
\thanks{Corresponding author. \href{mailto:mogavero@math.unipd.it}{mogavero@math.unipd.it}}
\affiliation{Dipartimento di Matematica “Tullio Levi-Civita”, Università degli Studi di Padova \\ 
Via Trieste 63, 35121 Padova, Italy}
\affiliation{IMCCE, CNRS, Observatoire de Paris, Universit\'{e} PSL, Sorbonne Universit\'{e} \\ 
77 Avenue Denfert-Rochereau, 75014 Paris, France}
\author{Jacques Laskar}
\email{jacques.laskar@obspm.fr}
\affiliation{IMCCE, CNRS, Observatoire de Paris, Universit\'{e} PSL, Sorbonne Universit\'{e} \\ 
77 Avenue Denfert-Rochereau, 75014 Paris, France}


\date{\today}

\begin{abstract}
Extrasolar planetary systems commonly exhibit planets on eccentric orbits, with many systems located near or within mean-motion resonances, showcasing a wide diversity of orbital architectures. Such complex systems challenge traditional secular theories, which are limited to first-order approximations in planetary masses or rely on expansions in orbital elements—eccentricities, inclinations, and semi-major axis ratios—that are subject to convergence issues, especially in highly eccentric, inclined, or tightly-packed systems. To overcome these limitations, we develop a numerical approach to second-order perturbation theory based on the Lie transform formalism. Our method avoids the need for expansions in orbital elements, ensuring broader applicability and more robust convergence. We first outline the Hamiltonian framework for the 3-body planetary problem, and apply a canonical transformation to eliminate fast angle dependencies, deriving the secular Hamiltonian up to second order in the mass ratio. We then use the fast Fourier transform algorithm to numerically simulate, in an accurate way, the long-term evolution of planetary systems near or away from mean-motion resonances. Finally, we validate our methods against well-known planetary configurations, such as the Sun-Jupiter-Saturn system, as well as to exoplanetary systems like WASP-148, TIC 279401253 and GJ 876, demonstrating the applicability of our models across a wide range of planetary configurations.
\end{abstract}
\maketitle
\section{\label{sec:intro}Introduction}
The quest to understand the long-term evolution of planetary systems dates back to the pioneering works of Laplace and Lagrange, who developed the first secular theories to predict planetary motion over extended timescales. These theories introduced the concept of secular dynamics, wherein short-term variations caused by the stellar attraction are averaged out to isolate the slower, long-term orbital evolution, driven by mutual planetary interactions. The Laplace-Lagrange theory, developed within this framework, has been a cornerstone for studying the long-term evolution of planetary systems, particularly for the nearly circular and coplanar orbits typical of the solar system \citep{Lagrange1781,Lagrange1782,Laplace1773}. Le Verrier \citep{LeVerrier1855} and Boquet \citep{Boquet1889} later extended this model to higher degrees in eccentricities and inclinations, aiming for a more accurate description of the solar system’s evolution. However, these early efforts, being first-order approximations in planetary masses, were insufficient to account for perturbations arising from mean-motion resonances (MMRs) between the outer planets. To address these limitations, second-order secular theories were developed, introducing corrections at second order in planetary mass ratios, and thus incorporating the effects of near-resonances on the long-term evolution of planetary orbits \citep{Hill1882, Brouwer1950, Laskar1985}. Among these developments, Laskar’s work combined numerical integration with extensive analytical secular theories, assisted by computer algebra, to explore the solar system's evolution over extended timescales. This approach revealed the chaotic behavior of the solar system and highlighted the crucial role of second-order terms in the long-term dynamics \citep{Laskar1988,Laskar1989,Laskar1990}. These second-order terms are particularly important for accurately capturing the slow precession of the perihelia and nodes of the planetary orbits. Precession frequencies are dominated by first-order terms in planetary masses,  meaning that second-order terms introduce relative corrections of order one to these frequencies,  which can be significant, especially for systems near MMRs. For example, for the Jupiter-Saturn system, which lies near the 5:2 MMR known as the Great Inequality, a first-order secular theory  poorly reproduces the fundamental frequencies $g_5$ and $g_6$, which dominate the perihelion precession of Jupiter and Saturn, respectively. The relative errors in these frequencies are approximately $ 14 \%$ for $g_5$ and $20\%$ for $g_6$. However, when second-order terms are included, these errors are reduced to below $1\%$.

The discovery of numerous exoplanetary systems has since revealed a remarkable diversity of orbital configurations. Unlike the nearly circular and coplanar orbits of the solar system, many exoplanetary systems feature planets with large eccentricities and masses, such as HD 183263 and HD 80606, with the latter reaching an eccentricity of 0.93. Some systems, like Kepler-36 and Kepler-90, exhibit tightly packed, non-hierarchical orbits \citep{seager2011exoplanets,Winn2015exoplanets}. Many of these systems are also either near or locked in MMRs, adding further complexity to their dynamics \citep{kley2006exoplanets,Lissauer2011exoplanets}. This diversity has reignited the interest in secular theories as a means to better understand the long-term orbital evolution of these newly discovered systems, constrain their formation processes, and characterize their overall orbital architectures. At first order in planetary mass, multipole secular theories, based on expansions in the semi-major axis ratio, have been developed for systems exhibiting hierarchical structure of orbits \citep{Hansen1855,Tisserand1894,Kozai1962,Lidov1962,Ford2000,Lee2003,Laskar2010}. For systems away from MMRs, Hamiltonian expansions up to high degrees in eccentricities have been used to describe the secular evolution of coplanar 3-body exoplanetary systems \citep{Libert2005,Libert2006}. These expansions were later extended to second-order in planetary masses to accurately account for the dynamics of systems near or inside MMRs \citep{Libert2013,Sansottera2019}. Other analytical approaches have also been developed at first order in planetary masses to describe resonant interactions between two planets in coplanar orbits. These methods often rely on simplifying assumptions to facilitate the analytical treatment of resonant dynamics \cite{Beauge2003,Callegari2006,Batygin2013,Delisle2012}.

In addition to analytical methods, numerical averaging approaches have emerged to address the limitations of analytical expansions, particularly convergence issues arising from truncations. The earliest numerical method, introduced by Gauss \citep{Gauss1818old}, treats planets as gravitationally interacting Keplerian rings, allowing the computation of secular orbital evolution without relying on truncated series expansions \citep{Hill1882,Musen1970}. Similarly, Schubart's method uses numerical averaging to handle MMRs in the restricted 3-body problem, allowing for the precise treatment of asteroid dynamics and small bodies in resonant configurations \citep{Schubart1964,Moons1993,Moons1994}. However, both methods are first-order in planetary masses, which limits their applicability to systems where higher-order interactions are not significant.

In this paper, we present a numerical approach to second-order perturbation theory based on the Lie transform formalism. We consider the 3-body problem consisting of two planets orbiting a star (we will still routinely use the generic term \textit{$n$-body} to distinguish the original, non-averaged dynamics from the secular one). Unlike traditional analytical expansions, our numerical approach is free from the limitations associated with truncations in eccentricities, inclinations, or semi-major axis ratios. This allows us to effectively handle the broad spectrum of exoplanetary systems, including those with complex orbital configurations and resonant interactions. We begin by outlining the Hamiltonian formalism for the planetary 3-body problem and specify our choice of coordinates (Sect.~\ref{sec:nbody}). Next, we apply the Lie series approach to perform a canonical transformation that eliminates fast-angle dependencies in the $n$-body Hamiltonian, providing the expressions for first- and second-order secular Hamiltonians (Sect.~\ref{sec:2ndOrderAveraging}). We then use the fast Fourier transform (FFT) algorithm to numerically evaluate and integrate the secular equations of motion (Sect.~\ref{sec:numerics}). Finally, we apply our method to various planetary systems, comparing its accuracy against direct $n$-body simulations (Sect.~\ref{sec:Applications}).

\section{\texorpdfstring{\resizebox{!}{0.007\textheight}{$n$}}{n}-body Hamiltonian}
\label{sec:nbody}
By using canonical heliocentric variables \citep{Laskar1991}, the 3-body Hamiltonian reads  
\begin{equation}
\label{eq:Hamiltonian}
H = \underbrace{\sum_{i=1}^2 \left( \frac{\Vert \vec{\tilde{r}}_i \Vert^2}{2 \beta_i} 
- \frac{\mu_i \beta_i}{\Vert \vec{r}_i \Vert} \right)}_{H_0}
+ \underbrace{\vphantom{\sum_{i=1}^2} \frac{\vec{\tilde{r}}_1 \cdot \vec{\tilde{r}}_2}{m_0} - \frac{G m_1 m_2}{\Vert \vec{r}_1 - \vec{r}_2 \Vert}}_{R \, = \, \epsilon H_1},
\end{equation}
where $m_0$ is the mass of the star, $m_i \ll m_0$ are the planetary masses, $\vec{\Tilde{r}}_i$ and $\vec{r}_i$  are the barycentric 
momenta and heliocentric position vectors of the planets, respectively; $G$ is the gravitational constant, $\beta_i= m_0m_i/(m_0 + m_i)$ are the reduced masses, and $\mu_i = G(m_0 + m_i)$. Hereafter, the indices $i = 1, 2$ stand for the inner and outer planets, respectively. 

Away from any close encounter between the planets, the disturbing function $R$ is much smaller than the Keplerian part of 
the Hamiltonian $H_0$. Therefore, Eq.~\eqref{eq:Hamiltonian} is in quasi-integrable form and we write $R = \epsilon H_1$, with 
$\epsilon = \max(m_1,m_2)/m_0 \ll 1$. It is then natural to introduce canonical variables that trivially integrate the Hamiltonian $H_0$. 
Building on the Delaunay variables, we consider 
\begin{equation}
\label{eq:mod_delaunay}
\begin{aligned}
&\Lambda = \beta\sqrt{\mu a}, \quad &x = \sqrt{\Lambda} \sqrt{1 - \sqrt{1- e^2}} \, \E^{\iota \varpi}, \\
&\lambda = M + \varpi, \quad &y = \sqrt{2 \Lambda} \left(1- e^2\right)^{\frac{1}{4}} \sin(\inc/2) \E^{\iota \Omega}, 
\end{aligned}
\end{equation}
where $a$ represents the semi-major axis of a Kepler orbit, $e$ the eccentricity, $I$ the inclination, $\varpi$ the longitude of 
perihelion, $\Omega$ the longitude of the node, $\lambda$ the mean longitude, and $M$ the mean anomaly \citep{Laskar1991}. 
Throughout the paper, $\iota$ represents the imaginary unit and $\E$ the exponential operator. 
$(\Lambda_i,\lambda_i; x_i, -\iota \overline{x}_i; y_i, -\iota \overline{y}_i)_{i=1}^2$ are momentum-coordinate pairs of 
canonical variables, with the overbar denoting the conjugate of a complex variable. With this choice of variables, the Keplerian 
part of the Hamiltonian simply reads $H_0 = - \sum_{i=1}^{2} \mu_i^2 \beta_i^3 / 2 \Lambda_i^2$, 
and $(\Lambda_i,x_i,y_i)_{i=1}^2$ are integrals of motion for its flow. 

In the planetary problem, the mean longitudes $\lambda_i$ vary much faster than all the other canonical variables in  Eq.~\eqref{eq:mod_delaunay}, the timescale separation being set by $\epsilon^{-1}$. This separation of timescales allows us to  focus on the long-term evolution of the orbits by \textit{averaging out} the fast variables $\lambda_i$. In canonical perturbation theory, this elimination of fast angles is achieved through a near-identity canonical transformation of variables, which pushes the dependence on the fast variables in the Hamiltonian to higher orders of the planet-star mass ratio $\epsilon$, resulting 
in a so-called secular Hamiltonian. In our study, we aim to construct the secular Hamiltonian up to second order in the mass ratio. To achieve this, we employ the Lie transform formalism, which provides a systematic approach to construct the necessary canonical  transformation \citep{Hori1966,Deprit1969}. 

It is worth noting that the above averaging procedure is only meaningful in regions of phase space where there is no overlap of MMRs causing short-term chaos. Therefore, in this work, we implicitly assume that the short-term dynamics of the planetary system is regular, with no sensible diffusion of the mean-motion frequencies $\vec{n} = \nabla\!_{\vec{\Lambda}} H $. Secular Hamiltonians, in contrast, are perfectly suited to model long-term chaos arising from overlap of secular resonances, as in the inner solar system \citep{Laskar1990,Laskar1992,Mogavero2022}.

\section{Second-order secular Hamiltonians}
\label{sec:2ndOrderAveraging}
Let us denote the fast angle variables by $\vec{\lambda} = (\lambda_1,\lambda_2)$, with conjugate momenta 
$\vec{\Lambda} = (\Lambda_1,\Lambda_2)$, and the slow coordinates by $\vec{q}=-\iota(\bar{x}_1,\bar{y}_1,\bar{x}_2,\bar{y}_2)$, 
with conjugate momenta $\vec{p}= (x_1,y_1,x_2,y_2)$. With this notation, the Hamiltonian reads
\begin{equation}
H(\vec{p},\vec{q};\vec{\Lambda},\vec{\lambda}) = H_0(\vec{\Lambda}) + \epsilon H_1 (\vec{p} , \vec{q} ;  \vec{\Lambda} ,\vec{\lambda}).
\end{equation}

We then consider the canonical transformation of variables $\vec{p},\vec{q}, \vec{\Lambda},\vec{\lambda} \rightarrow \vec{\hat p},\vec{\hat q},\vec{\hat \Lambda},\vec{\hat \lambda}$ defined as the time-$1$ flow of a generating function $S=S(\vec{\hat p}, \vec{\hat q}; \vec{\hat \Lambda},\vec{\hat \lambda})$ to be determined, that is, 
\begin{equation}
\label{eq:coordTransform}
\vec X=  \E^{\L_S} \vec{\hat X}
\! = \! \sum_{n=0}^{\infty} \frac{1}{n!} \L_{S}^n \vec{\hat X},
\end{equation}
where $\vec{X}$ is any of the variables $\vec{p},\vec{q}, \vec{\Lambda},\vec{\lambda}$ and $\vec{\hat X}$ is the corresponding transformed variable. In Eq.~\eqref{eq:coordTransform}, $\L_S \cdot= \{S, \cdot \}$ is the Lie derivative operator associated to the generating function, the braces $\{\cdot,\cdot\}$ representing the Poisson bracket, and $\L_{S}^n = \L_S \L_{S}^{n-1}$ for $n \geq 1$ \cite{Morbidelli2002}.
We recall that the Poisson bracket is given by 
\begin{multline}
 \{f,g\} = \frac{\partial f}{\partial \vec{\hat p}} \cdot \frac{\partial g}{\partial \vec{\hat q}}
 - \frac{\partial f}{\partial \vec{\hat q}} \cdot \frac{\partial g}{\partial \vec{\hat p}} + \\
 \frac{\partial f}{\partial \vec{\hat \Lambda}} \cdot \frac{\partial g}{\partial \vec{\hat \lambda}}
- \frac{\partial f}{\partial \vec{\hat \lambda}} \cdot \frac{\partial g}{\partial \vec{\hat \Lambda}} ,
\end{multline}
where the central dot stands for the scalar product of vectors. Both the generating function $S$ and the Poisson bracket are invariant under the canonical transformation in Eq.~\eqref{eq:coordTransform}.

The transformed Hamiltonian reads
\begin{equation}
\label{eq:averaged_H}
\widehat{H}(\vec{\hat p}, \vec{\hat q}; \vec{\hat \Lambda},\vec{\hat \lambda}) 
= \E^{\L_S} H(\vec{\hat p}, \vec{\hat q}; \vec{\hat \Lambda},\vec{\hat \lambda}).
\end{equation}
By writing the generating function as a series in the mass ratio, i.e. $S = \sum_{n=1}^\infty S_n \epsilon^n$, and expanding 
Eq.~\eqref{eq:averaged_H} as a Lie series, one obtains $\widehat{H} = \sum_{n=0}^\infty \widehat{H}_n \epsilon^n$ with 
\begin{equation}
\label{eq:averaged_H_series}
\begin{aligned}
\widehat{H}_0 &= H_0,\\
\widehat{H}_1 &= H_1 + \{ S_1 , H_0 \}, \\
\widehat{H}_2 &= \{ S_1 , H_1 \} + \frac{1}{2} \{ S_1 , \{ S_1 , H_0 \} \} + \{ S_2 , H_0 \}.
\end{aligned}
\end{equation}
In these expressions, both the right and left hand terms are
written and evaluated in the new variables $\vec{\hat p},\vec{\hat q},\vec{\hat \Lambda},\vec{\hat \lambda}$. This series expansion allows us to systematically eliminate the fast angle dependence in the Hamiltonian at each order in $\epsilon$ by choosing the appropriate series expansion of the generating function. In what follows, we detail the averaging procedure for both non-resonant and resonant planetary systems.

\subsection{Non-resonant systems}
If the planets are not inside any MMR, the dependence on the mean longitudes $\vec{\lambda}$ can be formally removed at any order in the mass ratio.
\paragraph{First-order averaging.}
\label{sec:first_order}
At first order in $\epsilon$, the secular Hamiltonian is obtained by imposing 
\begin{equation}
\label{eq:homological_1st}
H_1 + \{ S_1 , H_0 \} \vcentcolon= \left<H_1\right>,
\end{equation}
where the angle brackets denote averaging over the fast angles $\vec{\hat \lambda}$, that is, 
\begin{equation}
\left< f \right> = \frac{1}{(2 \pi)^2} \int_{\mathbb{T}^2} f \, d\vec{\hat \lambda} \, ,
\end{equation}
with $f$ an arbitrary phase-space function and the integration extending over the torus $\mathbb{T}^2 = [0,2\pi) \times [0,2\pi)$ 
at fixed values of all the remaining canonical variables. Eq.~\eqref{eq:homological_1st} is commonly referred to as the 
homological equation and its solution 
is obtained by expanding $H_1$ and $S_1$ as Fourier series in the angles $\vec{\hat \lambda}$. By writing 
\begin{equation}
\label{eq:FouriesSeriesH1}
H_1 (\vec{\hat p}, \vec{\hat q}; \vec{\hat \Lambda},\vec{\hat \lambda}) =\sum_{\vec{k} \in \mathbb{Z}^2} h_1 ^{\vec{k}} (\vec{\hat p},\vec{\hat q}; \vec{\hat \Lambda}) \, \E^{\iota \vec{k}\cdot\vec{\hat \lambda}} , 
\end{equation}
the generating function at first order is then given by 
\begin{equation}
\label{eq:Generatingfct$s_1$}
S_1 (\vec{\hat p}, \vec{\hat q}; \vec{\hat \Lambda},\vec{\hat \lambda})  = - \iota \sum_{\vec{k} \in \mathbb{Z}^2 \backslash \{\vec{0}\}} \frac{h_1^{\vec k} (\vec{\hat p}, \vec{\hat q}; \vec{\hat \Lambda})}{\vec k \cdot \vec{n}_0} \, \E^{\iota \vec{k}\cdot\vec{\hat \lambda}} ,
\end{equation}
where $\vec{n}_0 = \nabla\!_{\vec{\hat \Lambda}} H_0 $ is the vector of the planet mean-motions appearing in the solution of the integrable approximation $H_0$. 
As a result of Eq.~\eqref{eq:homological_1st}, at first order in the mass ratio, the secular Hamiltonian is simply the average of the $n$-body Hamiltonian over the fast angles $\vec{\hat \lambda}$, that is, 
\begin{equation}
\label{eq:secularHam_1}
\widehat{H} = H_0 + h^{\vec{0}}_1 \, \epsilon + O(\epsilon^2). 
\end{equation}
The dynamics generated by this Hamiltonian, known as Gauss dynamics \cite{Gauss1818old}, has been extensively used in both analytical and numerical studies, as discussed in the introduction.  For small planetary masses and away from any MMR, this approach effectively reproduces the long-term behaviour of the  planetary orbits. 

\paragraph{Second-order averaging.}
\label{sec:second_order}
When the planetary masses are not so small or planets are close to (but not in) a MMR, terms of order $\epsilon^2$ in 
Eq.~\eqref{eq:secularHam_1} can still raise a non-negligible contribution, resulting in a first-order approximation lacking precision when compared to the long-term behaviour of $n$-body dynamics. In this case, the averaging procedure has to be extended to second order in the mass ratio. 
By taking into account Eq.~\eqref{eq:homological_1st}, the second-order contribution to the Lie-transformed Hamiltonian in 
Eq.~\eqref{eq:averaged_H_series} becomes  
\begin{equation}
\widehat{H}_2 = \frac{1}{2} \{ S_1 , H_1 \} + \frac{1}{2}\{ S_1 , h_1^{\vec 0} \} + \{ S_2 , H_0 \}. 
\end{equation}
By noting that $\left< \{S_1, h_1^{\vec 0} \} \right> = 0$, one then solves a homological equation for $S_2$ requiring that $\widehat{H}_2$ is given by 
\begin{equation}
\widehat{H}_2 \vcentcolon= \frac{1}{2} \left<\{ S_1 , H_1 \} \right>.
\end{equation}
The term $\{S_1 , H_1\}$ is readily computed from Eqs.~\eqref{eq:FouriesSeriesH1} and \eqref{eq:Generatingfct$s_1$}: 
\begin{multline}
\{S_1,H_1\} = -\sum_{\vec{l} \in \mathbb{Z}^2} \E^{\iota \vec{l} \cdot \vec{\hat \lambda}} \Bigg[
\sum_{\vec{k} \in \mathbb{Z}^2 \backslash \{\vec{0}\}} \frac{\iota \{h_1^{\vec{k}}, h_1^{\vec{l}-\vec{k}} \}^{\boldsymbol{\star}} }{\vec{k}\cdot\vec{n}_0} \ + \\ 
(\vec{k}-\vec{l}) \cdot \frac{\partial}{\partial \vec{\hat \Lambda}} \biggl(\frac{h_1^{\vec{k}}}{\vec{k}\cdot\vec{n}_0}\biggr)  h_1^{\vec{l}-\vec{k}} 
+ \vec{k} \cdot \frac{\partial h_1^{\vec{l}-\vec{k}}}{\partial \vec{\hat \Lambda}} \frac{h_1^{\vec{k}}}{\vec{k}\cdot\vec{n}_0} \Bigg] ,
\end{multline}
where the star denotes the Poisson bracket restrained to the slow variables $\vec{\hat p},\vec{\hat q}$, that is,
\begin{equation}
\{f,g\}^{\boldsymbol{\star}} = \frac{\partial f}{\partial \vec{\hat p}} \cdot \frac{\partial g}{\partial \vec{\hat q}}
- \frac{\partial f}{\partial \vec{\hat q}} \cdot \frac{\partial g}{\partial \vec{\hat p}}. 
\end{equation}
After averaging over $\vec{\hat \lambda}$, one obtains the secular Hamiltonian at second order in the mass ratio as 
\begin{widetext}
\begin{equation}
\label{eq:secularHam}
\widehat{H} = H_0 + h^{\vec{0}}_1 \, \epsilon  - \frac{1}{2} \sum_{\vec{k} \in \mathbb{Z}^2 \backslash \{\vec{0}\}} \left[ \frac{\iota \{h_1^{\vec{k}}, \overline{h_1^{\vec{k}}} \}^{\boldsymbol{\star}}}{\vec{k}\cdot \vec{n}_0} + \vec{k}\cdot \frac{\partial}{\partial \vec{\hat \Lambda}} \biggl(\frac{|h_1^{\vec{k}}|^2}{\vec{k}\cdot \vec{n}_0}\biggr) \right] \epsilon^2 + O(\epsilon^3).
\end{equation}
\end{widetext}
The starred Poisson bracket arises from the fact that we only eliminated the fast angle variables $\vec{\hat \lambda}$. Eq.~\eqref{eq:secularHam} uses the 
property $h_1^{-\vec{k}} = \overline{h_1^{\vec{k}}}$, coming from the fact that the Hamiltonian is a real function.
The importance of second-order terms in Eq.~\eqref{eq:secularHam} is twofold. First, as the perihelion and nodal precession frequencies $\dot{\vec \varpi}$ and $\dot {\vec \Omega}$ are of order $\epsilon$, second-order terms introduce relative corrections of order $\epsilon$ to these frequencies, which can have a notable impact on the accuracy of the predicted long-term orbital evolution. Second, these terms involve divisors $\vec{k}\cdot \vec{n}_0$, which can be small near resonances, causing the second-order contributions to significantly influence the system's behavior even if the system is only close to, but not in MMR.

\paragraph{Mean-motion correction.}
\label{par:meanmotion}
As just mentioned, second-order terms are particularly sensitive to small divisors $\vec{k} \cdot \vec{n}_0$, which can significantly affect the system's dynamics near resonances. The accuracy of these divisors directly affects the ability of Eq.~\eqref{eq:secularHam} to precisely reproduce the secular frequencies $\dot{\vec \varpi}$ and $\dot {\vec \Omega}$, which govern the long-term evolution of planetary orbits. Ensuring that the frequencies $\vec{n}_0$ in these divisors closely match the actual mean-motion frequencies observed in the full $n$-body dynamics is essential for an accurate representation of the system's evolution. To achieve this, one can modify the unperturbed part of the Hamiltonian $H_0$ to incorporate key contributions from the perturbation $H_1$ that affect the mean-motion frequencies, while preserving the integrability of the unperturbed system. Following the approaches of \citep{Locatelli2000,Laskar1985}, we redefine the unperturbed Hamiltonian and the perturbation as follows:
\begin{equation}
\begin{aligned}
& H'_0 = H_0 + \epsilon \, h_1^{\vec 0} (\vec 0, \vec 0 ;\vec{\hat \Lambda}), \ \ H'_1 = H_1 - h_1^{\vec 0} (\vec 0, \vec 0 ;\vec{\hat \Lambda}), 
\end{aligned}
\end{equation}
where $\epsilon \, h_1^{\vec 0} (\vec 0, \vec 0 ;\vec{\hat \Lambda})$ is the $\vec{\hat \lambda}$-averaged disturbing function evaluated at zero eccentricity and inclination. With this adjustment, the mean-motion frequencies of the new unperturbed problem are given by 
\begin{equation}
\label{eq:modifiedmean}
\vec{n}'_0 = \nabla\!_{\vec{\hat \Lambda}} H'_0 = \vec{n}_0 + \epsilon \, \nabla\!_{\vec{\hat \Lambda}}  h_1^{\vec 0}(\vec 0, \vec 0 ;\vec{\hat \Lambda}).
\end{equation}
The frequencies $\vec{n}'_0$ provide a closer approximation to the $n$-body mean-motion frequencies, particularly in systems with low eccentricities and inclinations, where $h_1^{\vec 0} (\vec 0, \vec 0 ; \vec{\hat \Lambda})$ accounts for the dominant part of the perturbation. In systems with high eccentricities or inclinations, higher-degree terms in these variables also become important, but cannot be simply added to $H_0$ without disrupting the system’s integrability. The second-order secular Hamiltonian corresponding to the splitting $H = H'_0 + \epsilon H'_1$ is
\begin{widetext}
\begin{equation}
\label{eq:secularHam_correction}
\widehat{H} = H_0 + h^{\vec{0}}_1 \, \epsilon  - \frac{1}{2} \sum_{\vec{k} \in \mathbb{Z}^2 \backslash \{\vec{0}\}} \left[ \frac{\iota \{h_1^{\vec{k}}, \overline{h_1^{\vec{k}}} \}^{\boldsymbol{\star}}}{\vec{k}\cdot \vec{n}'_0} + \vec{k}\cdot \frac{\partial}{\partial \vec{\hat \Lambda}} \biggl(\frac{|h_1^{\vec{k}}|^2}{\vec{k}\cdot \vec{n}'_0}\biggr) \right] \epsilon^2 + O(\epsilon^3).
\end{equation}
\end{widetext}
We stress that the new splitting of the $n$-body Hamiltonian does not affect the expression previously derived for the secular Hamiltonian at second-order, except that the zeroth-order mean-motion vector $\vec{n}_0$ is replaced by the corrected vector $\vec{n}'_0$. Eq.~\eqref{eq:secularHam_correction} defines the \emph{(non-resonant) secular models} considered in Sects.~\ref{sec:numerics} and \ref{sec:Applications}. 

\paragraph{Secular equations of motion.}
The long-term evolution of the planetary orbits is governed by Hamilton's equations corresponding to the secular Hamiltonian given in Eq.~\eqref{eq:secularHam_correction}. The momenta $\vec{\hat \Lambda}$ are integrals of motion for its flow, and the time evolution of the orbits is determined by the equations 
\begin{equation}
\label{eq:EOMxy}
\frac{d \vec{\hat p}}{dt} = -\iota \frac{\partial \widehat{H}}{\partial \vec{\hat{q}}}.
\end{equation}
As these equations govern the dynamics of the Lie-transformed variables, the evolution of the original variables $\vec{p}, \vec{q}, \vec{\Lambda}, \vec{\lambda}$ is obtained using the Lie transforms introduced in Eq.~\eqref{eq:coordTransform}. 

\subsection{Resonant systems}
\paragraph{Averaging.}
\label{sec:MMR}
When planets are inside a MMR, that is, $\vec{\ell} \cdot \vec{n}'_0 = 0$ with a given $\vec{\ell} \in \mathbb{Z}^2 \backslash \{\vec{0}\}$, the corresponding denominators make Eq.~\eqref{eq:secularHam_correction} ill-posed. This is the so-called small divisors problem of perturbation theory. To address this, one modifies Eq.~\ref{eq:homological_1st} to retain, at first order in $\epsilon$, the Fourier harmonics of $H_1$ associated with the harmonic index $\vec{\ell}$ and its integer multiples. The modified generating function at first order is given by 
\begin{equation}
\label{eq:Generatingfct$s_1$_MMR}
S_1 (\vec{\hat p}, \vec{\hat q}; \vec{\hat \Lambda},\vec{\hat \lambda})  = - \iota \sum_{\vec{k} \in \mathbb{Z}^2 \backslash \mathcal{R}} \frac{h_1^{\vec k} (\vec{\hat p}, \vec{\hat q}; \vec{\hat \Lambda})}{\vec k \cdot \vec{n}'_0} \, \E^{\iota \vec{k}\cdot\vec{\hat \lambda}} ,
\end{equation}
where $\mathcal{R} = \{ m \vec{\ell} \, \big| \, m \in \mathbb{Z}\}$ represents the set of resonant harmonic indices and includes the null wave-vector. We stress that the resonant harmonics no longer enter the generating function. 
At second order in the mass ratio, the averaging process generates resonant harmonics through the Poisson bracket $\{S_1, H_1\}$, while the bracket $\{S_1, \widehat{H}_1\}$ only generates non-resonant terms. The \textit{resonant} secular Hamiltonian of order $\epsilon^2$ is defined by retaining the resonant Fourier harmonics while averaging out the non-resonant ones via an appropriate term $S_2$ in the generating function:
\begin{widetext}
\begin{subequations}
\label{eq:secularHam_MMR}
\begin{align}
\widehat{H} = H_0 + \epsilon \sum_{\vec{l} \in \mathcal{R}} h_1^{\vec{l}} \, \E^{\iota \vec{l}\cdot \vec{ \hat \lambda}} + \epsilon^2 \sum_{\vec{l} \in \mathcal{R}} \widehat{h}_2^{\vec{l}} \, \E^{\iota \vec{l}\cdot \vec{ \hat \lambda}} + O(\epsilon^3), \\
\widehat{h}_2^{\vec{l}} = - \frac{1}{2} 
\sum_{\vec{k} \in \mathbb{Z}^2 \backslash \mathcal{R}} \left[ \frac{\iota \{h_1^{\vec{k}}, h_1^{\vec{l}-\vec{k}} \}^{\boldsymbol{\star}} }{\vec{k}\cdot\vec{n'}_0} \ +  
\vec{k} \cdot \frac{\partial}{\partial \vec{\hat \Lambda}} \biggl(\frac{h_1^{\vec{k}} h_1^{\vec{l}-\vec{k}}}{\vec{k}\cdot\vec{n}'_0}\biggr) 
- \vec{l} \cdot \frac{\partial}{\partial \vec{\hat \Lambda}} \biggl(\frac{h_1^{\vec{k}}}{\vec{k}\cdot\vec{n}'_0}\biggr)  h_1^{\vec{l}-\vec{k}} \right], \label{eq:secularHam_MMR_2}
\end{align}
\end{subequations}
\end{widetext}
where $\widehat{h}_2^{\vec{l}}$ denote the Fourier harmonics of $\widehat{H}_2$. 
In the resonant case, the combination of angles $\vec{\ell} \cdot \vec{\hat \lambda}$ varies much slower than $\vec{\hat \lambda}$. 
The corresponding timescale is typically intermediate between the orbital periods and the secular timescales over which the slow canonical variables vary. Eq.~\eqref{eq:secularHam_MMR} takes this semi-slow timescale into account to describe the long-term dynamics of the planetary system. This equation defines the \emph{resonant secular models} considered in Sects.~\ref{sec:numerics} and \ref{sec:Applications}. 

\paragraph{Secular equations of motion and small divisors.}
\label{par:smalldivisors}
In the case of a MMR, the action variables $\vec{\hat \Lambda}$ are not integrals of motion of the secular dynamics. The long-term evolution of the planetary orbits is governed by additional 
equations of motion for the variables $\vec{\hat \Lambda}$ and $\vec{\hat \lambda}$: 
\begin{equation}
\begin{aligned}
\label{eq:EOMlamda}
\frac{d \vec{\hat p}}{dt} = -\iota \frac{\partial \widehat{H}}{\partial \vec{\hat{q}}}, \\
\frac{d \vec{\hat{\Lambda}}}{dt} = -\frac{\partial \widehat{H}}{\partial \vec {\hat \lambda}}, \quad \frac{d \vec{\hat{\lambda}}}{dt} = \frac{\partial \widehat{H}}{\partial \vec{\hat \Lambda}}.
\end{aligned}
\end{equation}

The time variation of the actions $\vec{\hat \Lambda}$ implies that the zeroth-order mean motions $\vec{n}'_0$, and thus the divisors appearing in Eq.~\eqref{eq:secularHam_MMR_2}, change during the orbit evolution. Although the resonant index $\vec{\ell}$ does not appear in the summation, if the orbit evolution gets sufficiently close to some higher-order resonances, certain divisors can become small enough to raise an instability while numerically integrating the equations of motion~\eqref{eq:EOMlamda}. There are a couple of ways to avoid this. 

The first possibility is to consider a variant of our perturbative approach in which the zeroth-order mean motions are constant throughout the evolution. This is achieved by Taylor expanding the integrable Hamiltonian $H_0'$ around a reference value $\vec{\hat{\Lambda}}_0$:
\begin{equation}
\label{eq:linearized_H0}
H_0’(\vec{\hat{\Lambda}}) = H_0’(\vec{\hat{\Lambda}}_0) + \delta\vec{\hat{\Lambda}} \cdot  \nabla\!_{\vec{\hat{\Lambda}}} H_0’ \bigg|_{\vec{\hat{\Lambda}}_0} + \delta H_0'(\vec{\hat{\Lambda}}),
\end{equation}
where $\delta\vec{\hat{\Lambda}} = \vec{\hat{\Lambda}} - \vec{\hat{\Lambda}}_0$ and $\delta H_0' = \mathcal{O}({\delta\vec {\hat{\Lambda}}}^2)$. One then considers the following splitting of the 
$n$-body Hamiltonian:
\begin{equation}
\begin{aligned}
\widetilde{H}_0 = H_0' - \delta H_0', \ \ \epsilon \widetilde{H}_1 = \epsilon H_1' + \delta H_0' , 
\end{aligned}
\end{equation}
with the mean-motion frequencies of the corresponding unperturbed problem given by
\begin{equation}
\label{eq:fixed_n0}
\widetilde{\vec{n}}_0 = \nabla\!_{\vec{\hat{\Lambda}}} H_0’ \bigg|_{\vec{\hat{\Lambda}}_0} .
\end{equation}
The second-order resonant secular Hamiltonian corresponding to the splitting $H = \widetilde{H}_0 + \epsilon \widetilde{H}_1$ is
\begin{widetext}
\begin{equation}
\begin{aligned}
\label{eq:secularHam_MMR_linearized}
\widehat{H} = H_0 + \epsilon \sum_{\vec{l} \in \mathcal{R}} h_1^{\vec{l}} \, \E^{\iota \vec{l}\cdot \vec{ \hat \lambda}} + \epsilon^2 \sum_{\vec{l} \in \mathcal{R}} \widehat{h}_2^{\vec{l}} \, \E^{\iota \vec{l}\cdot \vec{ \hat \lambda}} + O(\epsilon^3) , \\
\widehat{h}_2^{\vec{l}} = - \frac{1}{2} 
\sum_{\vec{k} \in \mathbb{Z}^2 \backslash \mathcal{R}} \frac{1}{\vec{k}\cdot\widetilde{\vec{n}}_0} \biggl[ \iota \{h_1^{\vec{k}}, h_1^{\vec{l}-\vec{k}} \}^{\boldsymbol{\star}} \ + \vec{k} \cdot \nabla\!_{\vec{\hat \Lambda}} \bigl( h_1^{\vec{k}} h_1^{\vec{l}-\vec{k}} \bigr) - \vec{l} \cdot \nabla\!_{\vec{\hat \Lambda}} \bigl( h_1^{\vec{k}} \bigr) h_1^{\vec{l}-\vec{k}} \biggl] \, . 
\end{aligned}
\end{equation}
\end{widetext}
The zeroth-order mean-motion frequencies no longer depend on the action variables $\vec{\hat \Lambda}$, so that the divisors $\vec{k} \cdot \widetilde{\vec{n}}_0$ are constant throughout the orbit evolution, thus preventing numerical instabilities that may arise from their time dependence. 
We stress that, if $\vec{\hat{\Lambda}}_0$ is chosen as the time average of $\vec{\hat{\Lambda}}$
along a given orbit evolution (see Sec.~\ref{sec:numerics}), one has the upper estimate $\delta\vec {\hat{\Lambda}} = \mathcal{O}(\sqrt{\epsilon} \vec{\hat{\Lambda}}_0)$ as a result of the resonant dynamics, so that $\delta H_0' = \mathcal{O}(\epsilon)$. 

A second way to avoid numerical instabilities in the integration of Eqs.~\eqref{eq:EOMlamda} is to exploit the exponential decay of the Fourier coefficients of the Hamiltonian \citep{Arnold1963,Morbidelli2002}: one can truncate the summation in Eq.~\eqref{eq:secularHam_MMR_2} in a way to neglect contributions from those high-order harmonics that generate very small divisors. This is the approach that we follow in the applications of this study (see Sects.~\ref{sec:numerics_truncation} and \ref{sec:Applications}).

\section{\label{sec:numerics}Numerical Implementation}
In this work, we do not use traditional analytical expansions in orbital elements, such as eccentricities, inclinations, or semi-major axis ratios, to avoid the convergence issues they raise. We set up a numerical algorithm to evaluate the secular Hamiltonian $\widehat{H}$ and its associated equations of motion \eqref{eq:EOMxy} and \eqref{eq:EOMlamda}. In this algorithm, our primary goal is to compute the key components appearing in the secular Hamiltonian $\widehat{H}$, that is the Fourier coefficients of the perturbation $H_1$ along with their partial derivatives with respect to the canonical variables $\vec x,\vec y ,\vec \Lambda,\vec \lambda$. This section outlines the key steps in the implementation, including the truncation of the Fourier series of $H_1$, the computation of its Fourier coefficients, and the evaluation of the right-hand side of the equations of motion up to second order in the mass ratio $\epsilon$.

\paragraph{Truncation of Fourier series.}
\label{sec:numerics_truncation}
The Fourier series of the Hamiltonian, Eq.~\eqref{eq:FouriesSeriesH1}, involves an infinite sum of harmonics indexed by $\vec{k} = (k_1, k_2) \in \mathbb{Z}^2$. In practice, this expansion must be truncated for numerical implementation. The Fourier coefficients of the perturbation $H_1$ decay exponentially with the order of the Fourier harmonic, that is, $|\vec{k}| = |k_1| + |k_2|$ \citep{Arnold1963,Morbidelli2002}. This allows us to express $H_1$ as 
\begin{equation}
\label{eq:Truncation}
H_1 =\sum_{\vec{k} \in \mathbb{Z}^2, |\vec{k}| \leq K} h_1 ^{\vec{k}} (\vec{\hat p},\vec{\hat q}; \vec{\hat \Lambda}) \, \E^{\iota \vec{k}\cdot\vec{\hat \lambda}} + O(\epsilon^2),
\end{equation}
where $K$ is a truncation limit chosen such that the remainder of the summation, which we ignore, is considered of order $\epsilon^2$ (we recall that the disturbing function is defined as $R=\epsilon H_1$).
Consequently, the summations required for the second-order terms appearing in Eqs.~\eqref{eq:secularHam_correction} and \eqref{eq:secularHam_MMR_2} are performed over the harmonic order $|\vec{k}|$ up to the truncation limit $K$. For resonant systems, the resonant set $\mathcal{R}$ must also adhere to this limit, which means that only harmonics with $|\vec{k}| \leq K$ are retained. 

In the case of a MMR, the retention of very high-order harmonics in the perturbation $H_1$ can lead to small divisors in Eq.~\eqref{eq:secularHam_MMR_2} that make the numerical integration of the equations of motion \eqref{eq:EOMlamda} challenging. When this happens, we simply further truncate the summation in Eq.~\eqref{eq:secularHam_MMR_2} to a lower order $K' < K$ such that no numerical instabilities are encountered during the integration.

\begin{figure}
    \centering
    \includegraphics[width=\linewidth]{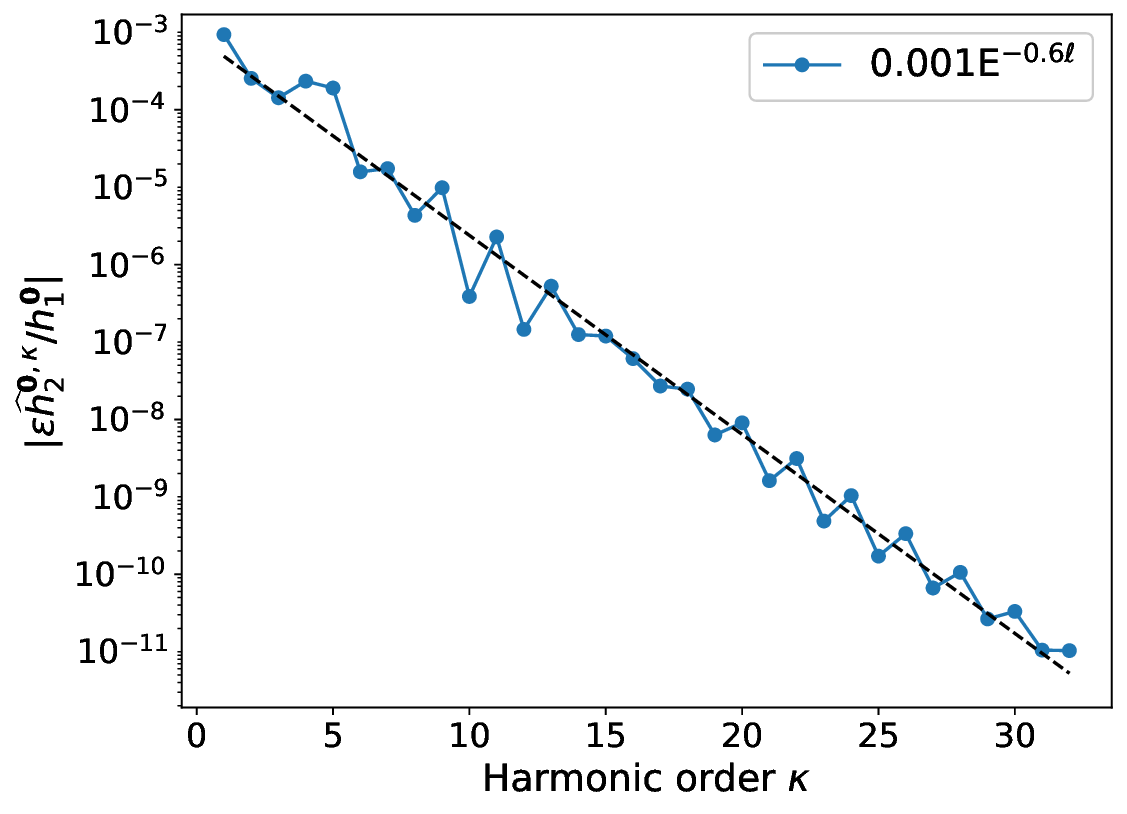}
    \caption{\textbf{Numerical convergence of $\widehat{H}_2$ for a non-resonant secular model of the Sun–Jupiter–Saturn system (see Eq.~\eqref{eq:secularHam_correction}).} Contribution at each harmonic order $\kappa$ to the zero-frequency harmonic $\epsilon^2 \widehat{h}_2^{\vec{0}}$ normalized by the first-order term $\epsilon h_1^{\vec 0}$, i.e., $|\epsilon \widehat{h}_2^{\vec{0}, \kappa}/h_1^{\vec{0}}|$, see Eq.~\eqref{eq:secondOrder}. The values are computed using the initial conditions of the Sun-Jupiter-Saturn system detailed in Sect.~\ref{app:JupSat}. The plot uses a log-linear scale, and the dashed line represents a fit, showing an exponential decrease of $\widehat{h}_2^{\vec{0},\kappa}$ with the harmonic order $\kappa$.}
    \label{fig:H2conv_sec}
\end{figure}

\paragraph{Convergence of second-order averaging.}
To assess the accuracy of our truncation scheme, we illustrate the convergence of the second-order term $\widehat{H}_2$ with the truncation limit $K$, taking the Sun–Jupiter–Saturn system as an example (see Sect.~\ref{app:JupSat}). We consider both non-resonant and resonant secular models, defined in Eqs.~\eqref{eq:secularHam_correction} and \eqref{eq:secularHam_MMR}, respectively.
In the non-resonant model, $\widehat{H}_2$ includes only the zero-frequency harmonic $\widehat{h}_2^{\vec{0}}$. We define the contribution to this term from a given harmonic order $\kappa \in \mathbb{N}^+$ as
\begin{equation}
\label{eq:secondOrder}
\widehat{h}_2^{\vec{0}, \kappa} = - \frac{1}{2} \sum_{\substack{\vec{k} \in \mathbb{Z}^2 \\ |\vec{k}|=\kappa}} \left[ \frac{\iota \{h_1^{\vec{k}}, \overline{h_1^{\vec{k}}} \}^{\boldsymbol{\star}}}{\vec{k}\cdot \vec{n}'_0} + \vec{k}\cdot \frac{\partial}{\partial \vec{\hat \Lambda}} \biggl(\frac{|h_1^{\vec{k}}|^2}{\vec{k}\cdot \vec{n}'_0}\biggr) \right]. 
\end{equation}
We then compute the normalized value $|\epsilon \widehat{h}_2^{\vec{0},\kappa}/h_1^{\vec{0}}|$ as a function of $\kappa$, at the initial conditions of the non-resonant model detailed in Sect.~\ref{app:JupSat}. The results are shown in Fig.~\ref{fig:H2conv_sec}, plotted on a log-linear scale.

\begin{figure}
    \centering
    \includegraphics[width=\linewidth]{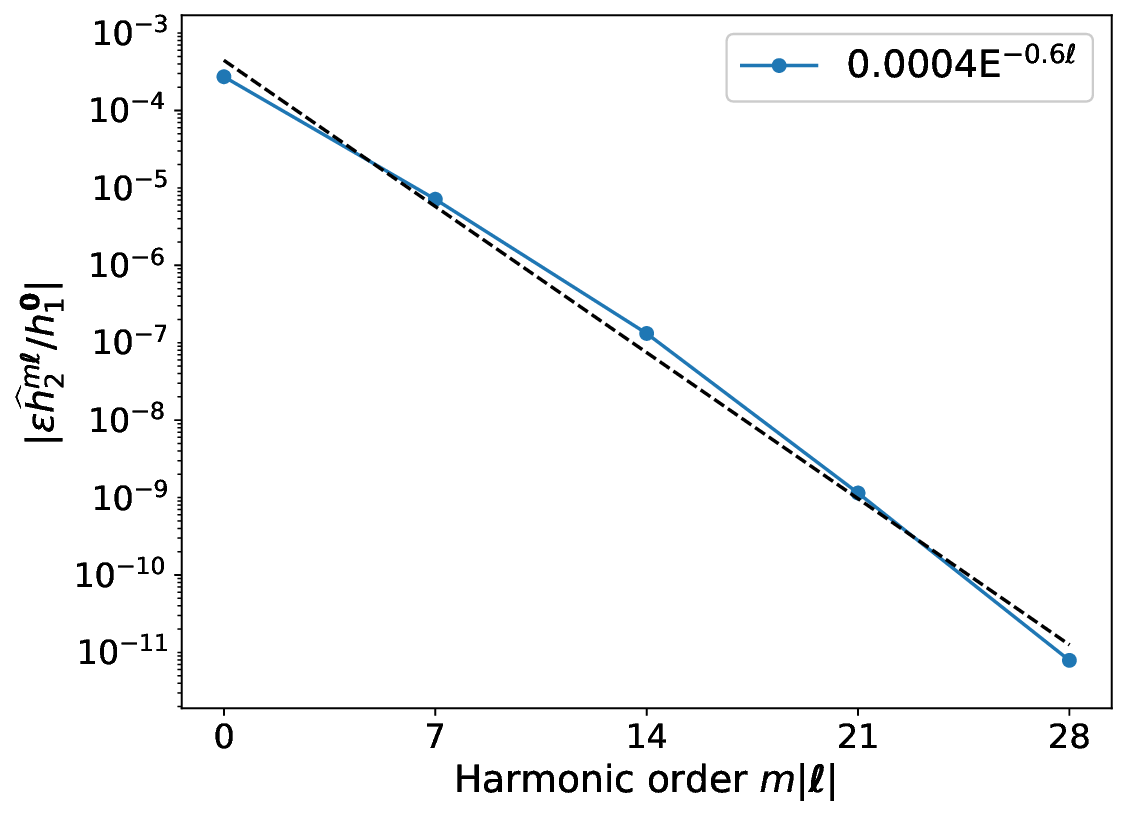}
    \caption{\textbf{Numerical convergence of $\widehat{H}_2$ for a resonant secular model of the Sun–Jupiter–Saturn system (see Eq.~\eqref{eq:secularHam_MMR}).} Fourier harmonics $\epsilon^2 \widehat{h}_2^{m \vec{\ell}}$ normalized by the first-order term $\epsilon h_1^{\vec{0}}$ and plotted as a function of the harmonic order $m |\vec{\ell}|$, where $\vec{\ell} = (-2, 5)$ corresponds to the 5:2 MMR between Jupiter and Saturn, and $m \in \mathbb{N}$. The plot uses a log-linear scale, and the dashed line represents a fit, showing an exponential decrease of $\widehat{h}_2^{m \vec{\ell}}$ with the harmonic order. }
    \label{fig:H2conv_res}
\end{figure}

For the resonant model, we evaluate the Fourier harmonics $\widehat{h}_2^{m \vec{\ell}}$ defined in Eq.~\eqref{eq:secularHam_MMR}, where $\vec{\ell} = (-2,5)$ corresponds to the 5:2 MMR between Jupiter and Saturn and $m \in \mathbb{N}$ indexes the resonant wave-vectors. The normalized values $|\epsilon \widehat{h}_2^{m \vec{\ell}} / h_1^{\vec{0}}|$ are computed at the initial conditions of the resonant model (see Sect.~\ref{app:JupSat}) and are shown in Fig.~\ref{fig:H2conv_res}, plotted on a log-linear scale.
The exponential decay observed in both non-resonant and resonant cases provides numerical evidence that the summations in Eqs.~\eqref{eq:secularHam_correction} and \eqref{eq:secularHam_MMR} are convergent and our second-order secular models are well-posed for the Sun–Jupiter–Saturn system.

\paragraph{Computation of Fourier coefficients via FFT.}
An essential step in the numerical implementation is the computation of the Fourier coefficients of $H_1$ and their partial derivatives with respect to the canonical variables. The computation of these derivatives is facilitated by the property that the derivatives of the Fourier coefficients of $H_1$ are the Fourier coefficients of the derivatives of $H_1$. Thus, we first derive closed-form analytical expressions for $H_1$ and its first-order partial derivatives as a function of the canonical variables. We then define a 2D grid for the angles $\lambda_1$ and $\lambda_2$, with $N$ points along each axis: 
\begin{equation}
\label{eq:lambdagrid}
\lambda_1,\lambda_2 \in \left\{ 2\pi k/N \, \big| \, k= 0,1,...,N-1 \right\}.
\end{equation}
On this grid, we evaluate $H_1$ and its partial derivatives and then perform a 2D fast Fourier transform \cite{Cooley1965} on the evaluated data to obtain the Fourier coefficients. The computational cost involves $N^2$ evaluations per each function, followed by a 2D FFT, with complexity $2 N^2 \log(N)$ per function. 
Note that performing an FFT with $N \times N$ points yields $N/2$ positive and $N/2$ negative frequency coefficients for each axis, so $N$ should always be chosen equal to $2K$ at least to ensure that we obtain the required number of Fourier coefficients.

\paragraph{Integration of equations of motion.}
\label{par:IntegrationsEOM}
With the FFT algorithm, we compute the Fourier spectrum of the Hamiltonian $H_1$ and its first-order partial derivatives with respect to the canonical variables. This spectrum is then used to evaluate the right-hand side of the secular equations of motion \eqref{eq:EOMxy} and \eqref{eq:EOMlamda} at first and second order in the mass ratio $\epsilon$.  At first order, the process is straightforward: we extract the zero-frequency harmonic and any harmonic within the resonant set $\mathcal{R}$, directly constructing the first-order secular Hamiltonian $\widehat H_1$ and the corresponding equations of motion. At second order, additional steps are required. Computing the second-order term $\widehat{H}_2$ in Eqs.~\eqref{eq:secularHam_correction} and \eqref{eq:secularHam_MMR} involves using all remaining Fourier coefficients of $H_1$ and their first-order derivatives that were not included at first order. These coefficients are readily available from the computed FFT spectra.
To incorporate second-order contributions in the equations of motion, we need to compute the partial derivatives of $\widehat{H}_2$ with respect to the canonical variables. This requires second-order derivatives of $H_1$, which are somewhat cumbersome to derive analytically. To address this, we compute the partial derivatives of $\widehat{H}_2$ using a numerical differentiation scheme, specifically the five-point stencil method, which is a fourth-order finite differences scheme \citep{abramowitz1965,sauer2012}. A discussion of the numerical error associated with this choice is presented in Sects.~\ref{sec:Applications} and \ref{sec:Conclusion}.

To integrate the equations of motion, we use an Adams PECE method of order 12 with a fixed time-step \cite{Laskar1994}. As we integrate secular equations, this time-step is typically much larger than the one used in direct $n$-body integrations. For each system we study, the time-step is chosen to be small enough to accurately resolve the fastest relevant timescale, while still allowing for efficient long-term integration. We stress that the Adams PECE method is well suited to our case where the evaluation of the equations of motion is costly.

\paragraph{Choice of initial conditions.}
\label{sec:initial_conditions}
Another crucial aspect of the numerical implementation is the choice of initial conditions for the Lie-transformed canonical variables $\vec{\hat p},\vec{\hat q},\vec{\hat \Lambda},\vec{\hat \lambda}$. Ideally, initial conditions should be computed by inverting the Lie transform in Eq.~\eqref{eq:coordTransform}. They should not be set to the initial value of the corresponding original variables, as this could cause an offset in the secular frequencies of the motion \citep{Laskar1987fitting}. Denoting as $\vec{X}$ any of the variables $\vec{p},\vec{q},\vec{\Lambda},\vec{\lambda}$ and as $\vec{\hat{X}}$ the corresponding transformed variable, Eq.~\eqref{eq:coordTransform} is readily inverted to give
\begin{widetext}
\begin{equation}
\label{eq:inverse_lie_transform}
\vec{\hat{X}} = \E^{-\L_S} \vec{X} = \vec{X} - \{S_1,\vec{X}\} \epsilon  + \left( \frac{1}{2} \{S_1,\{S_1,\vec{X}\} \} - \{S_2,\vec{X}\} \right) \epsilon^2 + O(\epsilon^3).
\end{equation}
\end{widetext} 
where the terms on the right-hand side are written and evaluated in the original variables (we recall that both the generating function $S$ and the Poisson bracket are invariant under the canonical transformation defined in Eq.~\eqref{eq:coordTransform}). Hereafter, $\vec{\hat{X}} \in \{  \vec{\hat p},\vec{\hat q},\vec{\hat \Lambda} \}$ will represent a secular, slow-varying variable. In the case of a MMR with index $\vec{\ell}$, this set will also includes the semi-slow variable $\vec{\ell} \cdot \vec{\hat \lambda}$. We then average the right-hand side of Eq.~\eqref{eq:inverse_lie_transform} over the fast angle variables $\vec \lambda$. This leads to the expression
\begin{equation}
\label{eq:var_transform}
\vec{\hat{X}} = \left<\vec{X} \right> +  \frac{1}{2} \left<\{S_1,\{S_1,\vec{X}\} \} \right> \epsilon^2 + O(\epsilon^3),
\end{equation}
where the term $S_2$ of the generating function no longer appears. Thus, up to order $\epsilon$, we have $\vec{\hat{X}} = \left<\vec{X}\right>$, meaning that the secular variables are simply the average of the original variables over the mean longitudes. 
In practice, such averaged variables can be computed numerically by filtering out the short-time components from the numerical solution of the full $n$-body equations of motion using a low-pass filter. Given the quasi-integrable nature of the Hamiltonian, the dynamical variables $\vec{X}$ exhibit a frequency spectrum with two distinct regimes: a high-frequency regime corresponding to the short-period orbital motions, and a low-frequency regime representing the slow secular variations of the orbits, with a frequency gap separating these regimes. In resonant systems, an additional semi-slow regime arises due to the resonant angles, which is also distinct from the high-frequency regime. This structure allows us to apply a low-pass filter that removes the high-frequency components while preserving the slow (and semi-slow) variations. A low-pass filter allows signals below a specified cutoff frequency to pass through while attenuating higher frequencies. Ideally, such a filter would completely eliminate frequencies above the cutoff, but practical filters have a gradual transition from the pass-band to the stop-band.

For our implementation, we selected a Butterworth filter due to its maximally flat frequency response in the pass-band, which minimizes oscillations in the retained signal. The design of this filter requires selecting both its order and the desired cutoff frequency. In general, high-order filters offer a steeper roll-off around the cutoff and provide better attenuation of unwanted frequencies, while being more subject to numerical instability. 
In our case, we use a 4th-order Butterworth filter, applied in both forward and backward directions, resulting in an 8th-order filter. This approach eliminates phase shifts typically introduced by such filters, while robustly attenuating high-frequency components. To maintain numerical stability, especially for small cutoff frequencies, we split the higher-order filter into second-order sections that are applied sequentially. This implementation is available in the standard Python library \textit{scipy} \cite{Scipy}, which we used. 
The cutoff frequency value is determined by analyzing the frequency spectrum of the variable $\vec X$: we perform a Fourier transform on the unfiltered $n$-body solution to identify the frequency gap between the short-period and slow secular components, then set the cutoff frequency within this gap. An example of this filtering procedure is given in Fig.~\ref{fig:x1Filtering} for the Sun-Jupiter-Saturn system studied in Sect.~\ref{sec:Applications}. 
After filtering, we set the initial conditions for the secular variables based on the filtered $n$-body solution. The specific choice depends on whether the system is modeled using non-resonant or resonant secular models, defined in Eqs.~\eqref{eq:secularHam_correction} and \eqref{eq:secularHam_MMR}, respectively.

For the non-resonant case, we assign the initial values of the secular variables $\vec{\hat{p}}$ and $\vec{\hat{q}}$ to the initial values of the filtered solution. Selecting initial conditions for the secular momenta $\vec{\hat \Lambda}$ requires particular care, as their values strongly influence, via the small divisors, the ability of the secular equations of motion \eqref{eq:EOMxy} to accurately reproduce the long-term evolution seen in $n$-body dynamics. We can set the initial values of $\vec{\hat \Lambda}$ to the mean of the filtered solution, since $\vec{\hat \Lambda}$ are constants of motion. 
Alternatively, we can fix the zeroth-order mean-motion vector $\vec n_0'$ in Eq.~\eqref{eq:modifiedmean} and solve for the momenta $\vec{\hat \Lambda}$. The values of $\vec n_0'$ can be directly  obtained from the filtered $n$-body solution. This follows from the Lie transform of the full mean-motion vector $\vec n = \vec{\dot{\lambda}} = \nabla\!_{\vec \Lambda } H_0 + \epsilon\nabla\!_{\vec \Lambda } H_1 $, which describes the time evolution of the mean longitudes $\vec \lambda$ in the original canonical variables. Using Eq.~\eqref{eq:coordTransform}, we can express the mean-motion vector in terms of the Lie-transformed variables as
\begin{equation} 
\begin{aligned}
\vec n = \vec{\dot{\lambda}}(\vec{p},\vec{q};\vec{\Lambda},\vec{\lambda}) 
= \E^{\L_S} \vec{\dot{\lambda}} (\vec{\hat p}, \vec{\hat q}; \vec{\hat \Lambda}, \vec{\hat \lambda}). 
\end{aligned}
\end{equation}
Expanding the Lie series in this expression and averaging over the fast angles $\vec{\hat \lambda}$, we obtain 
\begin{widetext}
\begin{equation}
\label{eq:Lie_meanmotion}
\left < \vec n \right> = \left <  \vec{\dot{\lambda}} \right> + \left( \left < \left \{ S_1, \frac{\partial H_1}{\partial \vec{\hat \Lambda}} \right\}\right>+\frac{1}{2} \left <\left \{ S_1, \left \{ S_1, \frac{\partial H_0}{\partial \vec{\hat \Lambda}} \right\} \right\}\right> \right) \epsilon^2  + \mathcal{O} (\epsilon^3) ,
\end{equation}
\end{widetext}
where the first term on the right-hand side is given by 
\begin{equation}
\begin{aligned}
\left <  \vec{\dot{\lambda}} \right> &=  \nabla\!_{\vec{\hat \Lambda}} H_0 + \epsilon \nabla\!_{\vec{\hat \Lambda}} h_1^{\vec 0}(\vec{ \hat p}, \vec{ \hat q} ;\vec{ \hat \Lambda}) \\
&= \vec{n}_0 + \epsilon \nabla\!_{\vec{\hat \Lambda}}  h_1^{\vec 0}(\vec 0, \vec 0 ;\vec{ \hat \Lambda})  + \epsilon \ \mathcal{O} ( e^2 , I^2) .
\end{aligned} 
\end{equation}
Here, we separated $h_1^{\vec 0}$ into two components: its contribution at zero eccentricity and inclination, and the remaining higher-degree terms in these variables. Using Eq.~\eqref{eq:modifiedmean}, we arrive at
\begin{equation}
\label{eq:meanmotion_correction}
\left<\vec n\right> = \vec{n}'_0 + \epsilon \ \mathcal{O} ( e^2 , I^2) + \mathcal{O} (\epsilon^2). 
\end{equation}
In practice, $\left< \vec n \right>$ corresponds to the filtered mean-motion obtained from the $n$-body solution. Therefore, to a first approximation, we can set the mean-motion vector $\vec{n}'_0$ equal to $\left< \vec n \right>$. By doing so, we neglect the quadratic and higher-degree terms in eccentricities and inclinations (this can be somewhat unsatisfactory for highly excited orbits). We as well discard $\epsilon^2$ corrections. In fact, in our approach of filtering an $n$-body solution to set the initial conditions of the secular model, we also neglect the $\epsilon^2$ term that appears in Eq.~\eqref{eq:var_transform}. These corrections can be computed numerically, but they are typically small and have minimal impact on the overall accuracy of the model. 
We stress that Eq.~\eqref{eq:meanmotion_correction} provides the rationale for the mean-motion correction we introduced in Sect.~\ref{par:meanmotion}, aimed at enhancing the overall accuracy of the secular model. 

For resonant systems, the initial conditions for the secular variables $\vec{\hat{p}}, \vec{\hat{q}}$ and $\vec{\hat{\Lambda}}$, as well as for the semi-slow variable $\vec{\ell} \cdot \vec{\hat{\lambda}}$, are assigned directly as the initial values of the filtered $n$-body solution. In cases where the integrable part of the Hamiltonian $\widetilde{H}_0$ is linearized to fix the mean-motion vector $\widetilde{\vec{n}}_0$ (see Sect.~\ref{par:smalldivisors}), a reference value $\vec{\hat{\Lambda}}_0$ must also be specified. In such a case, one can set $\vec{\hat{\Lambda}}_0$ as the mean value of $\vec{\Lambda}$ computed from the filtered $n$-body solution.
\section{Application to Planetary Systems}
\label{sec:Applications}
In this section, we apply our second-order secular models to various planetary systems to investigate their long-term orbital evolution. To validate the accuracy of our models, we benchmark the results we obtain against direct $n$-body integrations performed using the symplectic integrator \texttt{SABA1064} \citep{Farres2013}. For each system, we compare the time evolution of planetary eccentricities and the dominant secular frequencies obtained from our secular models with those extracted from the $n$-body integration. We note that, in all plots, the eccentricity curves from the secular models correspond to the Lie-transformed variables, whereas those from the $n$-body solutions are derived from the original canonical variables. Throughout this section, we refer to models without MMR terms as \textit{(non-resonant) secular models} (Eq.~\ref{eq:secularHam_correction}) and those with these terms as \textit{resonant secular models} (Eq.~\ref{eq:secularHam_MMR}).
\begin{figure*}[htp!]
\centering
\includegraphics[width=\linewidth]{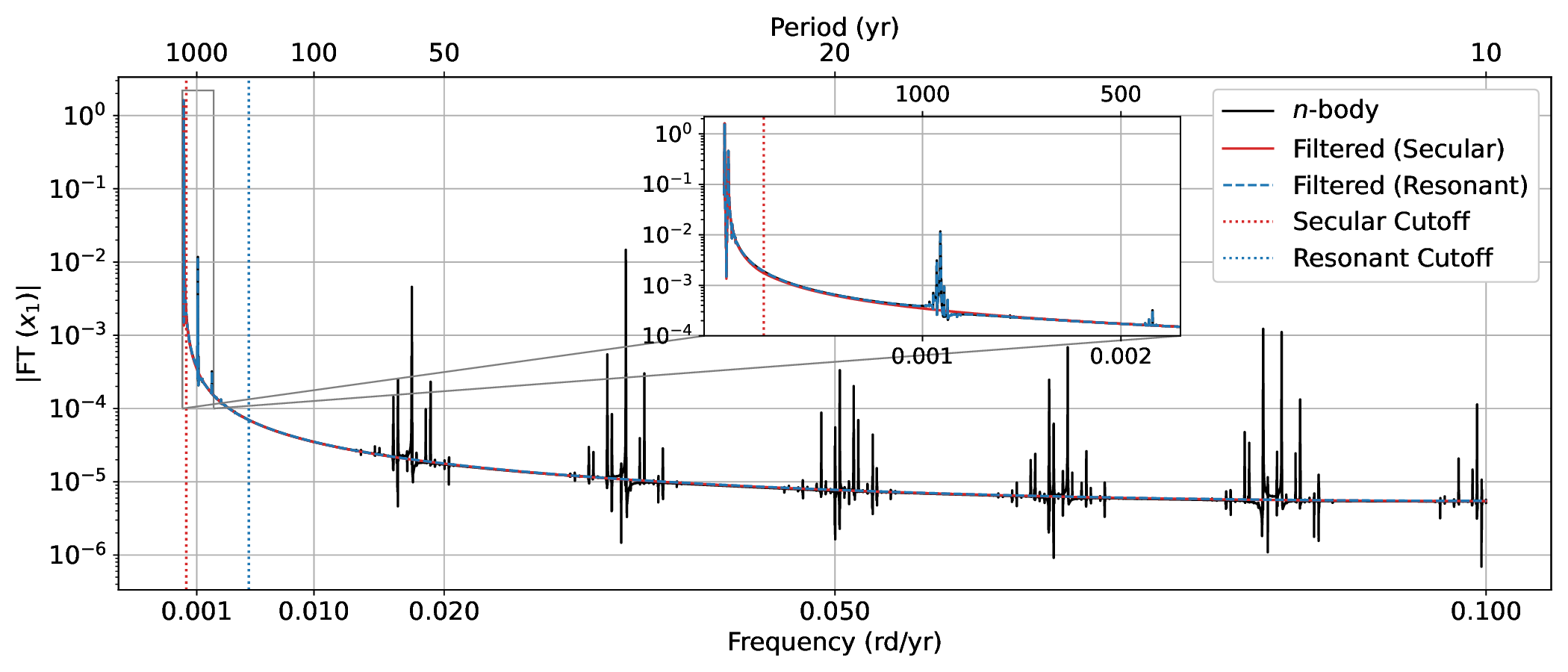} 
\caption{Amplitude of the Fourier transform of the variable  $x_1$ (proportional to Jupiter’s eccentricity $e_1$ at leading order) for the SJS system. The unfiltered $n$-body data (black) is shown alongside the data filtered using a Butterworth filter with cutoff periods of 5000 years (red) for the secular model and 200 years (blue) for the resonant secular model. The vertical dotted lines indicate the cutoff frequencies. The inset plot provides a zoomed-in view of the lower frequency range, clearly showing the effect of filtering in both cases to isolate the desired timescales.}
\label{fig:x1Filtering}
\end{figure*}
\paragraph*{Jupiter-Saturn Great Inequality.}
\label{app:JupSat}
The Jupiter-Saturn system, located near a 5:2 mean-motion resonance known as the Great Inequality, has been extensively studied both within the full solar system context and through focused analysis on a Sun-Jupiter-Saturn model (hereafter SJS). In the solar system context, the effort to develop an accurate secular theory for planetary orbital evolution dates back to Lagrange and was later extended by others. Early models included only first-order terms in planetary masses \citep{LeVerrier1855,Boquet1889}. Subsequent advancements introduced second-order corrections and incorporated the main contributions of the Jupiter-Saturn pair into their expansions, thereby ensuring a more precise representation of the planets' long-term dynamics \citep{Hill1897, Brouwer1950, Duriez1977, Laskar1985, Laskar1990}. In studies focused specifically on the SJS system, numerical methods have been utilized to investigate potential chaotic behaviors linked to the 5:2 MMR \citep{Varadi1999}. Analytical approaches, such as those by Michtchenko et al. \cite{Michtchenko2001b}, mapped the phase space near this resonance, while Locatelli and Giorgilli \citep{Locatelli2000, Locatelli2007} applied Kolmogorov-Arnold-Moser (KAM) theory to examine the system’s stability. In this study, we revisit the SJS system using both non-resonant and resonant second-order secular models to examine its long-term evolution. To setup the $n$-body model, we use planetary masses and initial orbital elements for Jupiter and Saturn extracted from the direct numerical integration LaX13b used in \cite{Mogavero2021}. 

To accurately select initial conditions for integrating the secular equations of motion, we apply the low-pass filtering approach described in the previous section.  For the (non-resonant) secular model, we select a cutoff frequency corresponding to a period of 5000 years \citep{Carpino1987}, which effectively removes short-term oscillations while preserving only the secular variations associated with the fundamental secular frequencies $g_5$, $g_6$ and $s_6$. For the resonant secular model, a higher cutoff frequency corresponding to a period of 200 years is used to ensure that resonant variations related to the 5:2 MMR, with a period of approximately 900 years, are preserved. Fig.~\ref{fig:x1Filtering} shows the amplitude of the Fourier transform for the canonical variable $x_1$, related to Jupiter's eccentricity, in both the secular and resonant secular models, demonstrating how low-pass filtering isolates the desired secular and resonant timescales in each model.
\begin{figure*}[htp]
\includegraphics[width=\columnwidth]{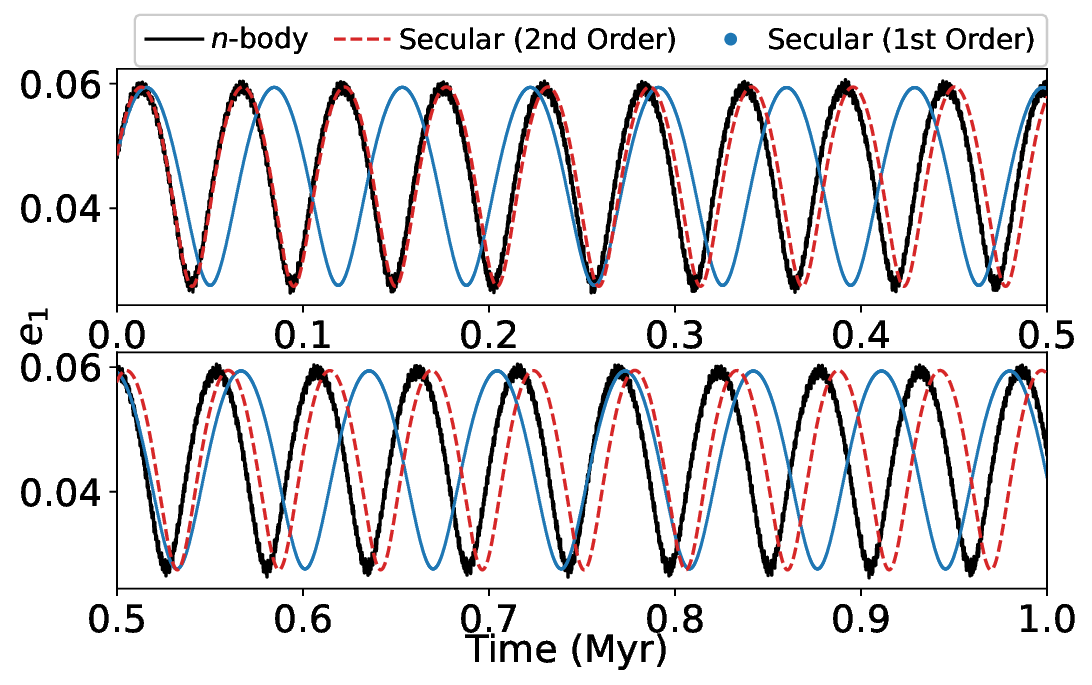} \includegraphics[width=\columnwidth]{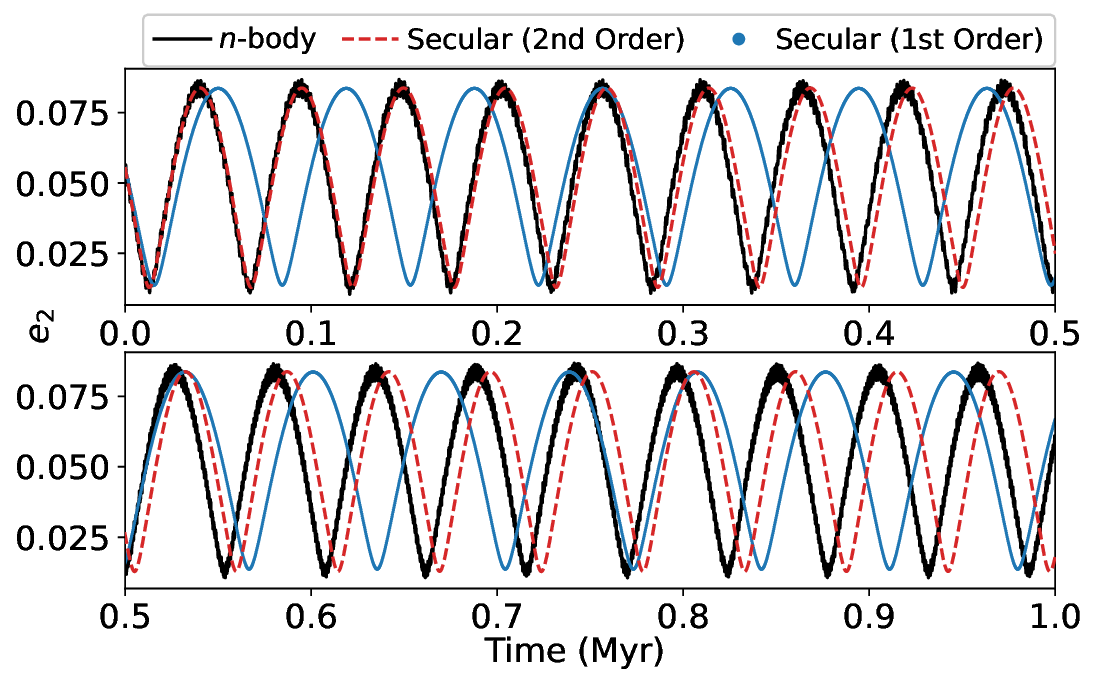} \\
\includegraphics[width=\columnwidth]{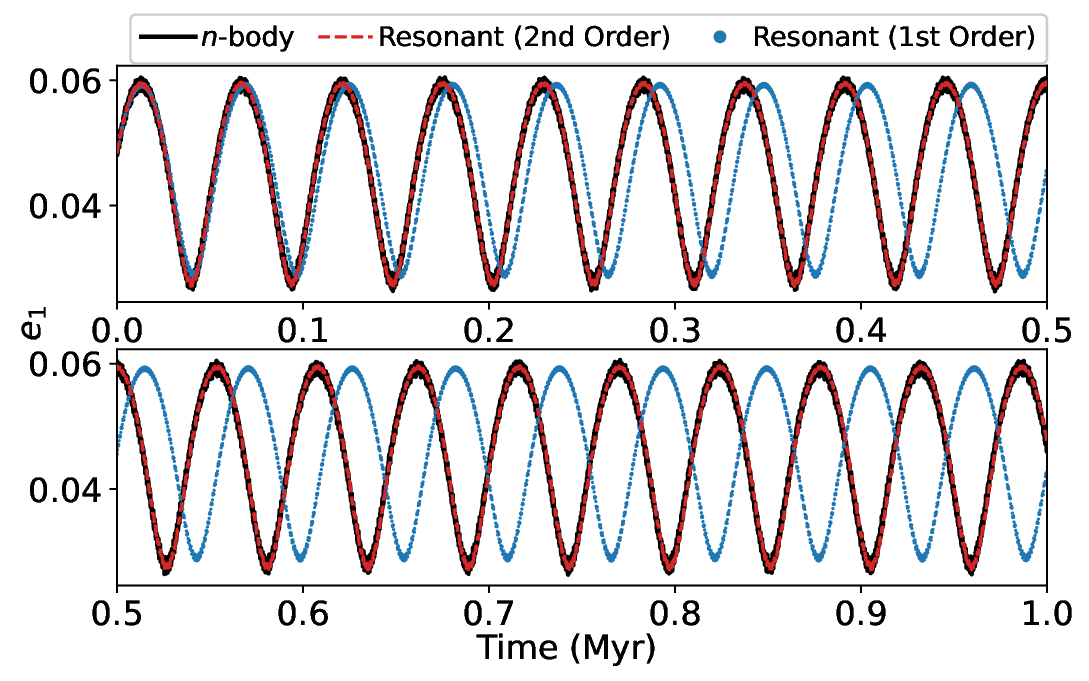} \includegraphics[width=\columnwidth]{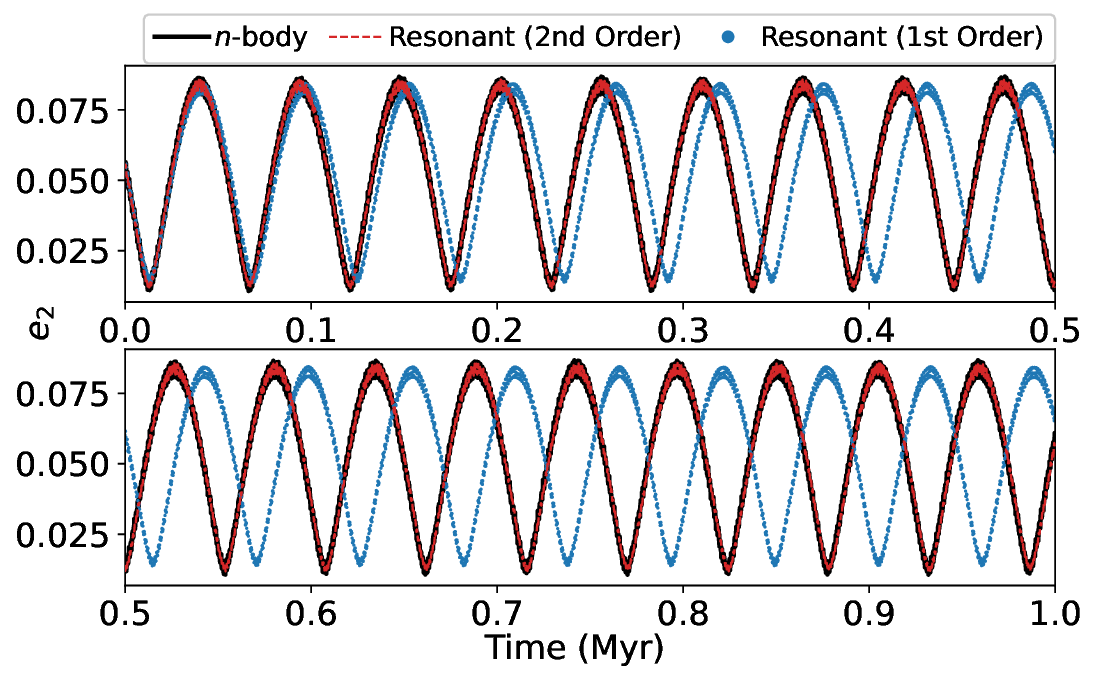}
\caption{Eccentricity evolution of Jupiter ($e_1$) and Saturn ($e_2$) over $1$ million years. The top row compares $n$-body results (black) with first-order (blue) and second-order (red) secular models. The bottom row shows similar comparisons for the resonant secular models.}
\label{fig:JSevolutionSec}
\end{figure*}

The FFT calculations required to compute the equations of motion were performed with a grid size of $64 \times 64$ points, and a truncation limit of $K=32$ was chosen for the Fourier expansion of $H_1$. To integrate the equations of motion, we use a time-step of 250 years when evolving the secular model. For the resonant secular model, a smaller time-step of 18 years is used to accurately capture the effects of the 5:2 near MMR. The integration error is monitored by tracking the conservation of energy and angular momentum throughout the integration. For example, in a 1 Gyr integration of the system, the relative error in energy remains on the order of $10^{-16}$ while that of the angular momentum components ranged between $10^{-15} - 10^{-13}$, for the first- and second-order models respectively. The slight increase in error for the second-order model is attributed to the use of a numerical derivative scheme for the derivatives of $H_2$ (see Sect.~\ref{par:IntegrationsEOM}), which introduces a minor impact on integration accuracy. To estimate the numerical derivative error, we computed the first-order partial derivatives of $H_1$ with respect to  the  variables  $\vec x$ and $\vec y$ using both analytical expressions and the numerical scheme. Across the phase-space region spanned by the secular model, the standard deviation of the relative error between the analytical and numerical derivatives ranged from $10^{-11}$ to $10^{-10}$, depending on the specific partial derivative considered. Since this scheme is applied only to terms of order $\epsilon^2$, it does not introduce significant discrepancies in the overall results.

Fig.~\ref{fig:JSevolutionSec} presents the evolution of the eccentricities of Jupiter and Saturn over a period of 1 million years, comparing our secular and resonant secular models with direct $n$-body simulations. The top row illustrates the results for the secular model, while the bottom row shows the resonant secular one. From the plots, it is clear that the first-order secular model captures the amplitudes of the eccentricity oscillations reasonably well, but fails to accurately reproduce the secular frequencies of motion when compared to direct $n$-body simulation, as shown by the phase shift of the curves. The inclusion of second-order terms improves the accuracy of these frequencies, though a gradual shift is still observed over longer integration times. When the 5:2 MMR terms are included in the secular Hamiltonian, the first-order model still misses the correct secular frequencies and performs no better than the second-order secular model. On the other hand, the second-order resonant secular model effectively reproduces both the amplitudes and frequencies of the motion, making it the most accurate approach among those tested.
To better quantify the performance of our models and precisely measure the differences in the secular frequencies, we performed frequency analysis \citep{Laskar1988,Laskar2005} on the complex variables $\vec z = \vec e \E^{\iota \vec \varpi}$ and $\vec \zeta = \sin{(\vec I /2)} \E^{\iota \vec \Omega}$ over a $[0,1]$ Myr interval for each orbital solution. The resulting secular frequencies $g_5$, $g_6$ and $s_6$  are presented  in Table~\ref{tab:secular_freq} for each model.
At first order, the secular theory yields significant relative errors in the secular frequencies — $14 \%$ for $g_5$ and $20 \%$ for $g_6$. Including the 5:2 MMR terms reduces these errors to  $11.5 \%$ on $g_5$  and  $4 \%$ on $g_6$, though this improvement remains insufficient for accurately modeling long-term evolution.
With the second-order secular model, errors reduce to  $0.3 \%$ for $g_5$ and $1 \%$ for $g_6$. Laskar \citep{Laskar1988} also noted this $0.28^{\prime\prime}/$yr difference in $g_6$ in his second-order model for the solar system, attributing it to higher-order terms in planetary masses for the outer planets. With the second-order resonant secular model, the relative error in $g_6$ is further reduced to $0.05\%$ confirming that the observed discrepancy in the secular theory mainly arises from higher-order terms in the planetary masses associated with the 5:2 MMR. 
\paragraph*{\label{app:Wasp148}WASP-148 system.}
Discovered by Hebrard et. al \citep{Hebrard2020}, the Wasp-148 system consists of two giant planets orbiting a G-type star with a mass of $1 \ M_{\odot}$, on eccentric and possibly inclined orbits. The planets have masses of $0.31 \ M_{\text{Jup}}$ and $0.41 \ M_{\text{Jup}}$ with eccentricities of $0.21$ and $0.18$, respectively. The inner planet, WASP-148b, has an orbital period of $8.80$ days, while the outer planet, WASP-148c, orbits in $34.54$ days, suggesting that these planets are near a 4:1 MMR. This proximity to resonance was confirmed by dynamical stability studies \citep{Hebrard2020, Almenara_2022}, which also estimated a mutual orbital inclination of about $20^{o}$. Here, we apply our first- and second-order secular models (Eqs.~\ref{eq:secularHam_1} and \ref{eq:secularHam_correction}, respectively) to investigate the secular evolution of this system, using the orbital parameters reported by~\citep{Almenara_2022}. As in the SJS  system, we perform the FFT calculations on a $64 \times 64$ grid, and a truncation limit of $K = 32$ was chosen for the Fourier expansion of $H_1$. The corresponding equations of motion were integrated using a fixed time-step of 1 year. 
\begin{table*}[htp]
\centering
\resizebox{0.75\textwidth}{!}{%
\begin{tabular}{r|c|c|c|c}
\hline\hline
\diagbox{Model}{System}  & SJS & WASP-148  & TIC 279401253 & GJ 876 \\ \hline 
&\begin{tabular}{c@{\hskip 0.25in}c@{\hskip 0.32in}c@{\hskip 0.09in}}
$g_5$ & $g_6$ & $s_6$ 
\end{tabular}
& \begin{tabular}{@{\hskip 0.2in}c@{\hskip 0.4in}c@{\hskip 0.4in}c@{\hskip 0.2in}}
$g_1$ & $g_2$& $s_2$ 
\end{tabular} 
&\begin{tabular}{@{\hskip 0.05in}c@{\hskip 0.35in}c}
$g_1$ & $g_2$ 
\end{tabular} 
&\begin{tabular}{@{\hskip 0.05in}c@{\hskip 0.35in}c}
$g_1$ & $g_2$ 
\end{tabular} \\
$n$-body & \begin{tabular}{r r r}
4.028 & 28.001 & -26.040
\end{tabular} 
& \begin{tabular}{c c c}
428.285 & 1182.647 & -2104.984
\end{tabular}&
\begin{tabular}{r  r}
-0.0659 & 0.0998
\end{tabular}&
\begin{tabular}{r  r}
-0.8807 & {\color{white}-}0.0202
\end{tabular} \\
Secular 1st & \begin{tabular}{r r r}
 3.484 & 22.301 & -26.061
\end{tabular}
&\begin{tabular}{r r r}
376.148 &  1106.215 &  -2102.458
\end{tabular}  
& \begin{tabular}{r  r}
\end{tabular}\\
Resonant 1st  &
\begin{tabular}{r r r}
3.559 & 26.809 & -26.085
\end{tabular}
& \begin{tabular}{r r r}
425.455  &  1176.596&  -2103.082
\end{tabular} &
\begin{tabular}{r  r}
-0.0658 & 0.0947
\end{tabular} &
\begin{tabular}{r  r}
-0.8882 & -0.0025
\end{tabular} \\
Secular 2nd & \begin{tabular}{r r r}
4.016 & 27.719 & -26.037
\end{tabular} 
&\begin{tabular}{r r r}
428.205 &  1182.190 & -2104.908
\end{tabular} &
\begin{tabular}{r  r}
    & 
\end{tabular}\\ 
Resonant 2nd &
\begin{tabular}{r r r}
4.026 & 27.986 & -26.042
\end{tabular}
&\begin{tabular}{r r r}
428.275 & 1182.650 & -2104.955
\end{tabular} &
\begin{tabular}{r  r}
-0.0666 & 0.1001
\end{tabular}&
\begin{tabular}{r  r}
-0.8832 & {\color{white}-}0.0197
\end{tabular} \\ \hline
\end{tabular}}
\caption{Comparison of secular frequencies for each planetary system using different models. Frequencies are reported in arcsec/yr for the SJS and WASP-148 systems, and in rd/yr for TIC and GJ 876 systems. The values are obtained through frequency analysis of $n$-body simulations and both first- and second-order secular and resonant secular models. For the TIC and GJ 876, we do not report the results of the first-order secular model as it fails to reproduce the systems' dynamics (see Fig.~\ref{fig:TICevolutionSec}); the second-order secular model is not considered as both systems are locked inside a 2:1 MMR.}
\label{tab:secular_freq}
\end{table*}
\begin{figure*}[htp!]
\includegraphics[width=\columnwidth]{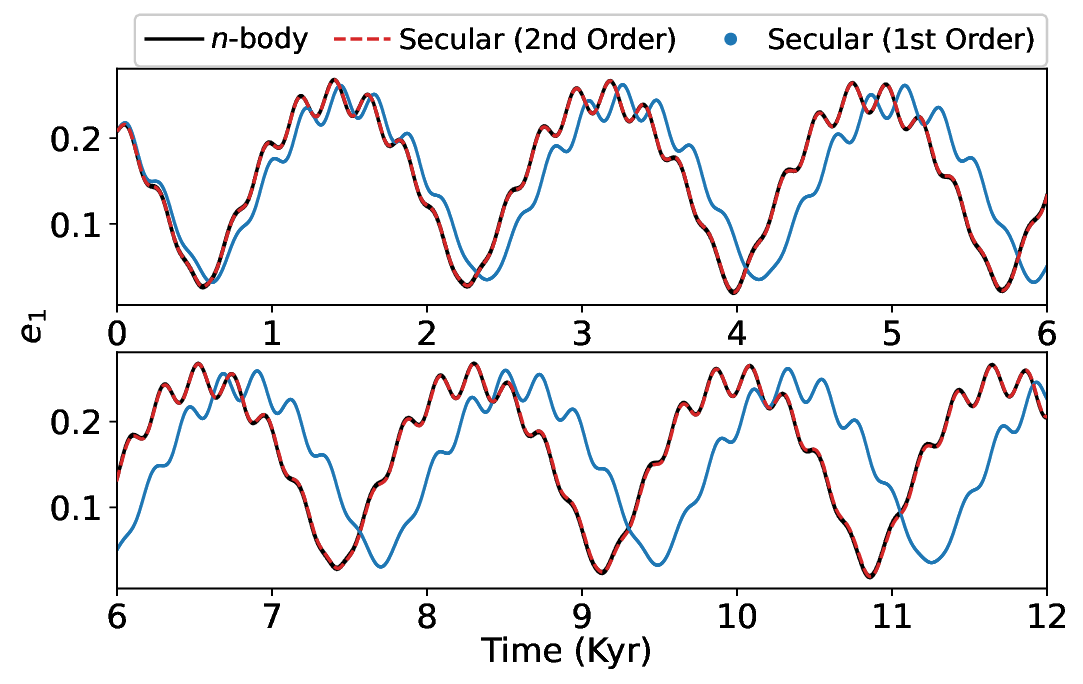} \includegraphics[width=\columnwidth]{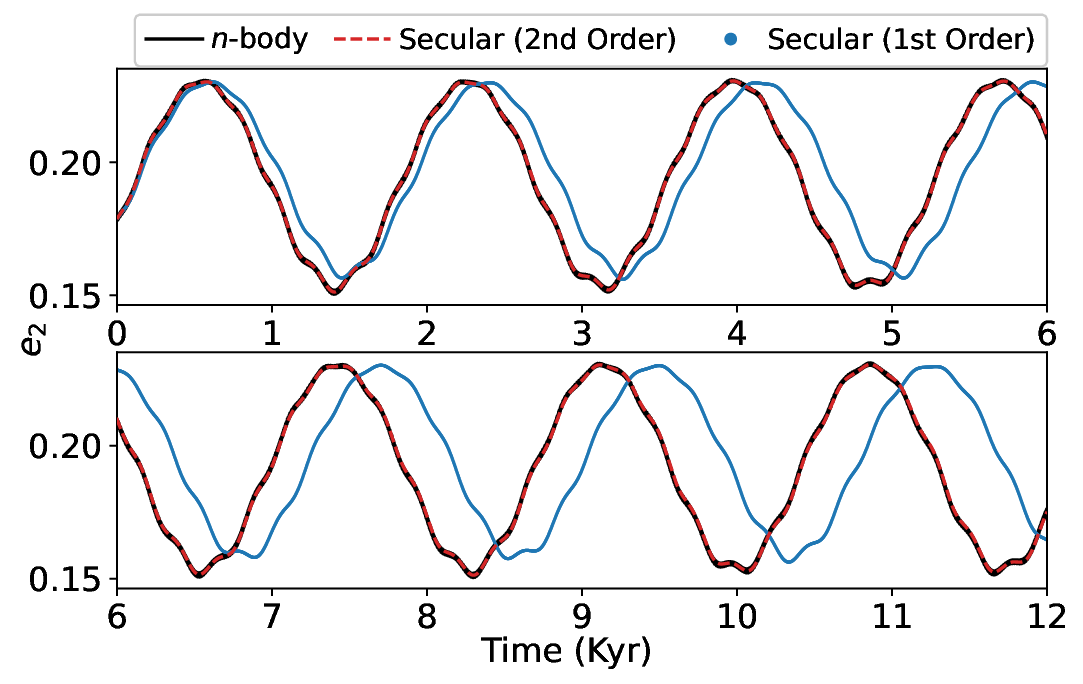}
\caption{Eccentricity evolution of the inner planet ($e_1$) and outer planet ($e_2$) in the WASP-148 system obtained by first-order secular model (blue), second-order secular model (red) and direct $n$-body integrations (black).}
\label{fig:waspevolution}
\end{figure*}
Fig.~\ref{fig:waspevolution} displays the eccentricity evolution for planets b ($e_1$) and c ($e_2$). We can see again that the first-order model captures the overall oscillation patterns but lacks the precision needed to reproduce the motion frequencies. The second-order model corrects this discrepancy and closely matches the $n$-body results. The secular frequencies $g_1$, $g_2$ and $s_2$ related to the perihelia and nodal precession are shown in Table~\ref{tab:secular_freq} for each model.
With the second-order model, relative errors reduce to below $0.1 \%$ for  $g_1$ and $g_2$, compared to $12 \%$ and $6.5 \%$ for the first-order model. As for the inclination components, the first-order model already captures $s_2$ well with an error of $0.1 \%$, which further reduces with the second-order terms. For completeness, we also simulate the system's evolution using first- and second-order resonant secular models. The characteristic period of the 4:1 MMR is approximately 1.2 years, and the equations of motion are integrated with a fixed timestep of $7.3$ days. At first order, the resonant secular model predicts secular frequencies with relative errors below $ 1 \%$, though it doesn't match the precision of the second-order secular approach. The second-order resonant secular model achieves the highest accuracy, with errors below $0.002\%$ for all frequencies.
\begin{figure*}[htp!]
\includegraphics[width=\columnwidth]{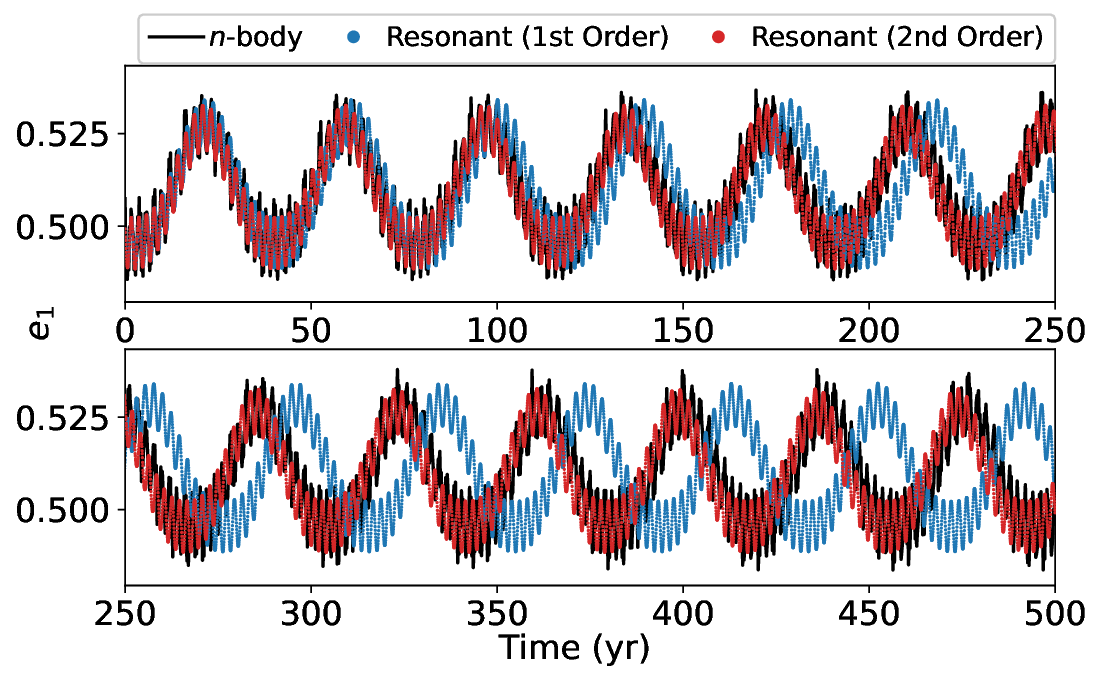} \includegraphics[width=\columnwidth]{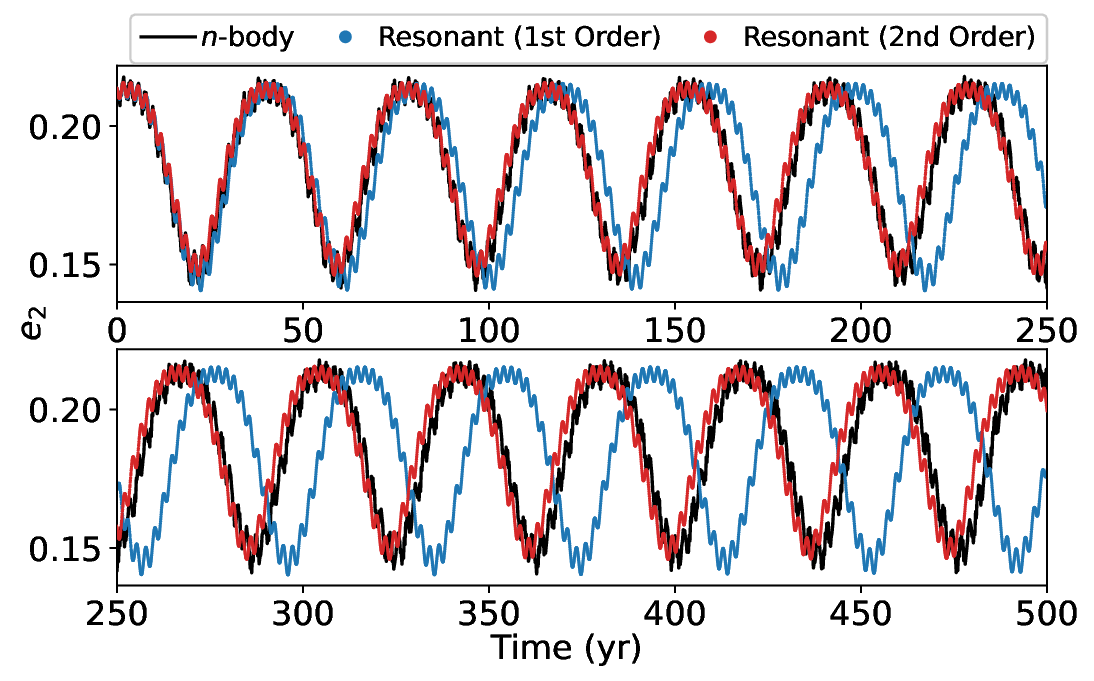}
\caption{Eccentricity evolution of the inner planet ($e_1$) and outer planet ($e_2$) in the TIC 279401253 system obtained by first-order resonant secular model (blue), second-order resonant secular model (red) and direct $n$-body integrations (black).}
\label{fig:TICevolution}
\end{figure*}
\begin{figure*}[htp!]
\includegraphics[width=\columnwidth]{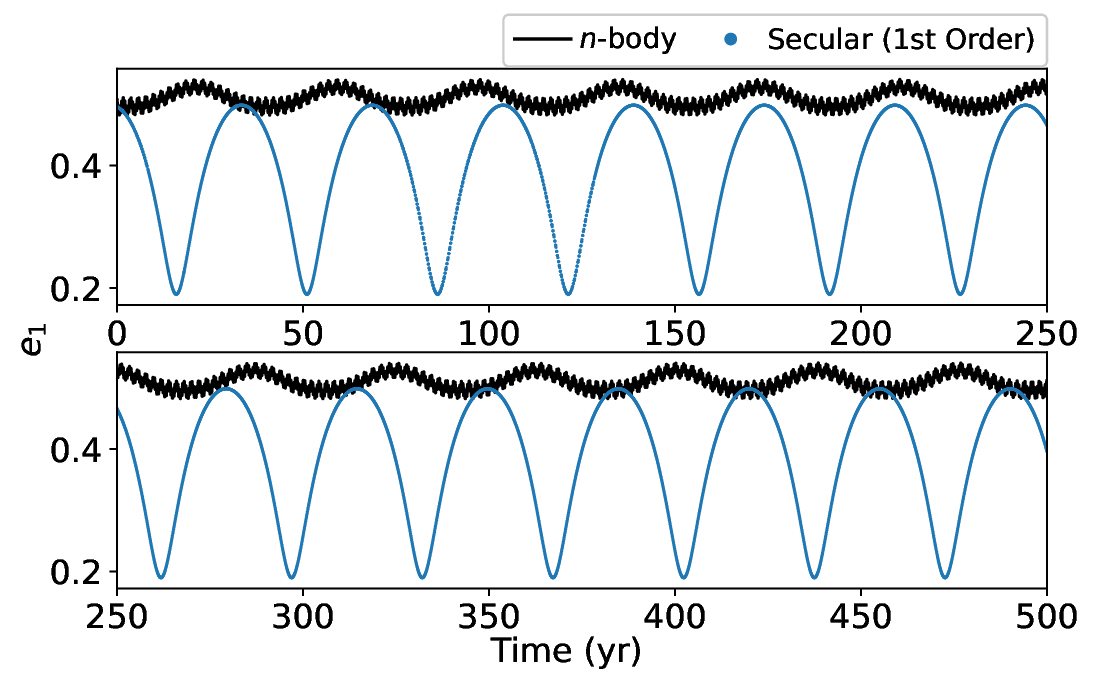} \includegraphics[width=\columnwidth]{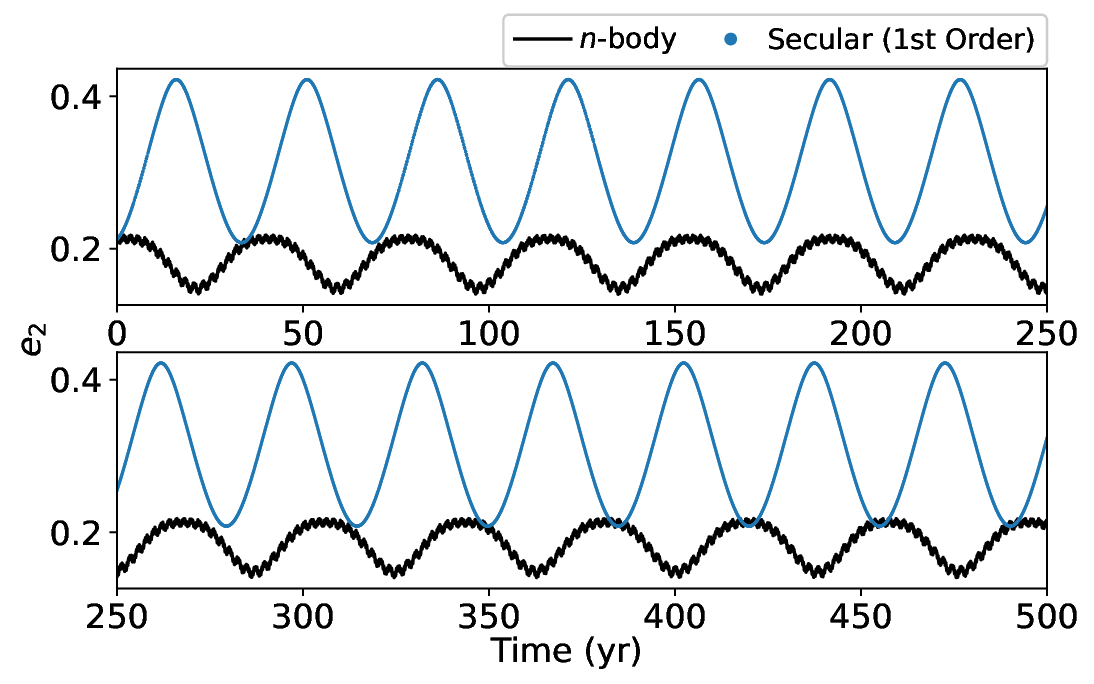}
\caption{Eccentricity evolution of the inner planet ($e_1$) and outer planet ($e_2$) in the TIC 279401253 system obtained by first-order (non-resonant) secular model (blue) and direct $n$-body integrations (black). As the system is inside MMR, this model fails to reproduce the system’s dynamics.}
\label{fig:TICevolutionSec}
\end{figure*}

\begin{figure}[htp!]
    \centering
    \includegraphics[width=\linewidth]{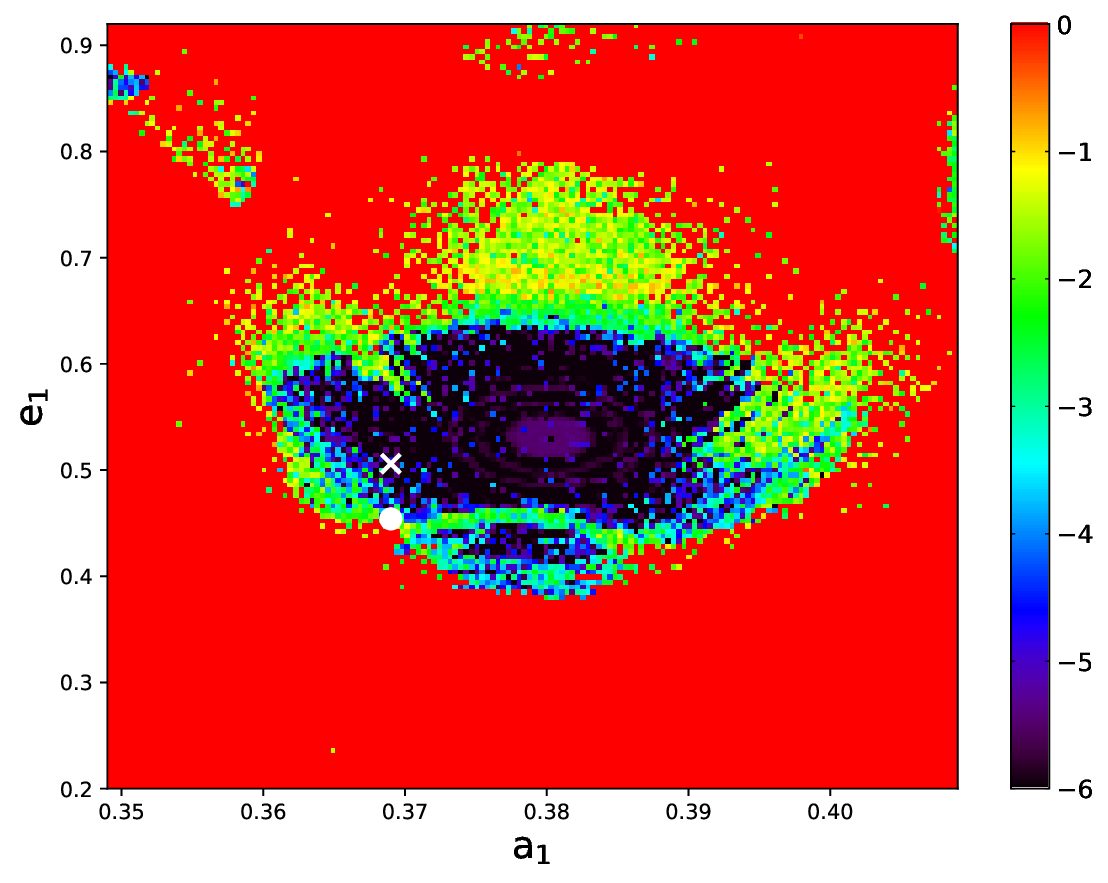}
    \caption{\textbf{Stability Analysis for the TIC 279401253 planetary system.} The phase space around the nominal solution from \citep{Bozhilov2023} is explored by varying the semi-major axis $a_1$ and eccentricity $e_1$ of the inner planet, while all other orbital elements are fixed. For each initial condition, the system is integrated over 10 kyr and frequency analysis of the mean longitude is performed. Chaotic diffusion is measured by the variation in the computed frequencies. The color scale shows the decimal logarithm of this indicator: dark blue denotes highly stable trajectories, while red indicates strongly chaotic motion. The white circle marks the nominal orbital configuration, while the white cross shows the modified solution used in our simulations, located deeper within the stable region.}
    \label{fig:TICmap}
\end{figure}

\paragraph*{\label{app:TIC}TIC 279401253 system.}
The TIC 279401253 system (hereafter TIC), reported in 2023 \citep{Bozhilov2023}, includes two massive planets orbiting a G-type star with a mass of $1.13 \ M_{\odot}$. The planets, with masses of $6.38 \ M_{\text{Jup}}$ and $8.12 \ M_{\text{Jup}}$, follow nearly coplanar, highly eccentric orbits, with eccentricities of $0.45$ and $0.22$, and orbital periods of $76.82$ days and $155.25$ days. Dynamical analysis of the system reveals that the planets are locked in a strong 2:1 MMR, with a resonant libration period of 2 years. Very interestingly, the nominal orbital solution from \citep{Bozhilov2023} places the system exactly at the boundary between chaotic and quasi-periodic short-term motions. This is illustrated in Fig.~\ref{fig:TICmap}, which shows a dynamical stability map of the TIC system in the $(a_1, e_1)$ plane—that is, the semi-major axis and eccentricity of the inner planet—constructed using frequency analysis of the mean longitudes \citep{Laskar1990, Laskar1993, Couetdic2010}. The stability indicator is reported using a color scale, with red indicating chaotic trajectories and dark blue denoting stable regions, that is, predictable short-term motion. The nominal solution, depicted as a white circle, lies at the edge of the stable zone, very close to the region of chaotic motion. To ensure a meaningful application of our secular models (see Sect.~\ref{sec:nbody}), we slightly increase the eccentricity of planet b—within its 2$\sigma$ observational uncertainty—so that the system lies deeper inside the stable region. This modified configuration is marked by the white cross.
We then apply our first-order and second-order resonant secular models, defined in Eq.~\eqref{eq:secularHam_MMR}, to explore the long-term evolution of the orbits. In this system, the convergence of the Fourier series of the perturbation $H_1$ is slower than in the previous cases, due to the higher eccentricities and planetary masses. Therefore, the FFT calculations required to compute the equations of motion were performed on a $128 \times 128$ grid of points to account for the contributions of higher-order harmonics. A truncation limit of $K = 64$ was then chosen for the Fourier expansion of the perturbation $H_1$. However, to avoid numerical instabilities related to small divisors in the second-order terms, the summation in Eq.~\eqref{eq:secularHam_MMR_2} was further truncated to $K' = 20$ (see Sect.~\ref{sec:numerics_truncation}). The corresponding equations of motion are integrated using a fixed time-step of $3.6$ days.
 
The evolution of the eccentricities for the TIC system over $500$ years is shown in Fig.~\ref{fig:TICevolution}. As with 
previous systems, the first-order Hamiltonian provides a good approximation but lacks precision in capturing the frequencies of motion. Including second-order terms in the Hamiltonian clearly reduces these discrepancies, though a gradual shift still appears over extended integration periods. The resultant secular frequencies $g_1$ and $g_2$ are reported in Table~\ref{tab:secular_freq} for each model.  At first order, the relative errors on $g_1$ and $g_2$ are $0.2\%$ and $5.1\%$, respectively. At second order, the error on $g_2$ drops to $0.3\%$, while the error on $g_1$ increases to $1\%$, resulting in an overall improvement in the accuracy of the secular model. It should be noted that the high planet eccentricities in this system probably make the mean-motion correction introduced in Sect.~\ref{par:meanmotion} less effective in this particular case. For completeness, we also simulate the system's evolution using the first-order (non-resonant) secular model. As shown in Fig.~\ref{fig:TICevolutionSec}, this model fails to reproduce the system’s dynamics, yielding oscillation patterns that differ significantly from the $n$-body results. We stress that this discrepancy is not due to the choice of initial conditions, which were derived by filtering an $n$-body solution as in all other cases; it rather reflects the limitation of the first-order secular theory in accurately modeling this system. We also recall that, as the system is locked inside a 2:1 MMR, one should not apply the second-order secular model in Eq.~\eqref{eq:secularHam_correction} due to the small divisors problem.

\paragraph*{\label{app:GJ876}GJ 876 system.}
The GJ 876 system is a well-known multi-planetary system consisting of four planets orbiting an M-type star with a mass of $0.34 \ M_{\odot}$ \citep{Marcy2001}. It was one of the first exoplanetary systems discovered to exhibit a resonant chain configuration, where successive planet pairs are locked in MMRs. In particular, planets GJ 876 b and GJ 876 c are inside a 2:1 MMR, leading to strong mutual gravitational interactions and rapid variations in their orbital elements. The libration period of the associated resonant angle is approximately 1.2 years. This makes the system a key benchmark for studies of resonant dynamics. In this study, we consider a two-planet model that includes only planets b and c, and we adopt masses and orbital elements reported in \citep{Laughlin2001}. Planet c has an orbital period of $29.91$ days and a mass of $0.92 \ M_{\text{Jup}}$, while planet b orbits with a period of $60.31$ days and a mass of $3.08 \ M_{\text{Jup}}$. The two planets follow coplanar orbits, with eccentricities of $0.252$ and $0.046$, respectively. We apply our first-order and second-order resonant secular models, defined in Eqs.~\eqref{eq:secularHam_MMR}, to study the long-term evolution of the pair. The FFT calculations required to evaluate the equations of motion were performed on a $128 \times 128$ grid of points. A truncation limit of $K = 64$ was used for the Fourier expansion of the perturbation $H_1$, while the summation in Eq.~\eqref{eq:secularHam_MMR_2} was further restricted to $K' = 20$ to avoid small divisor-related instabilities as in the TIC system. The equations of motion are then integrated using a fixed time-step of 1.8 days.
\begin{figure*}[htp!]
\includegraphics[width=\columnwidth]{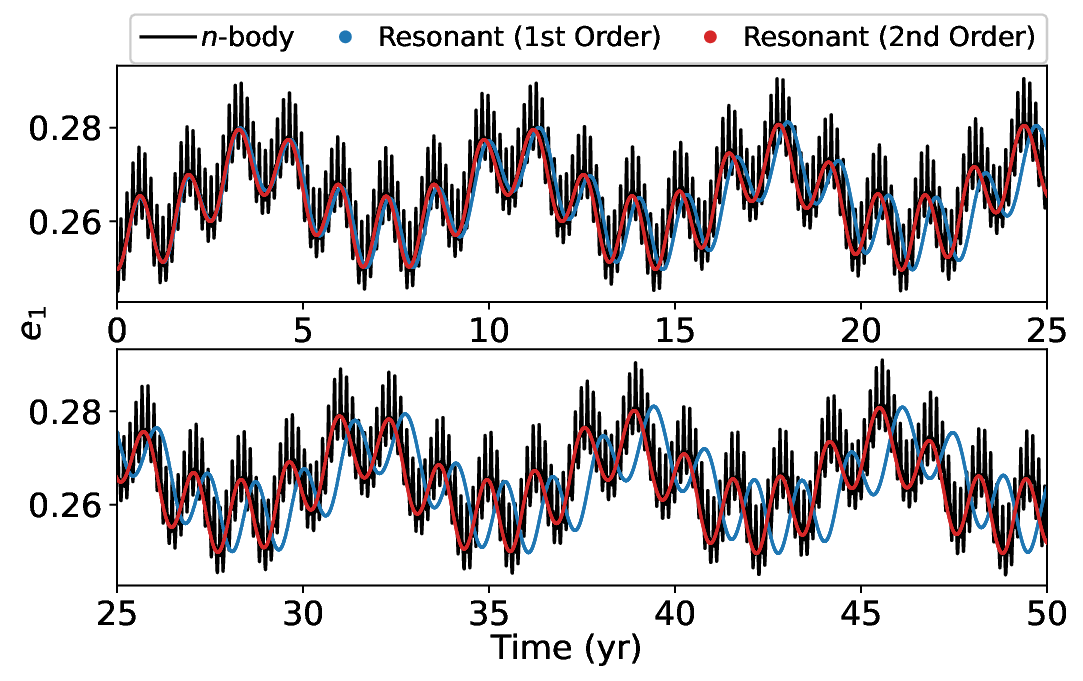} \includegraphics[width=\columnwidth]{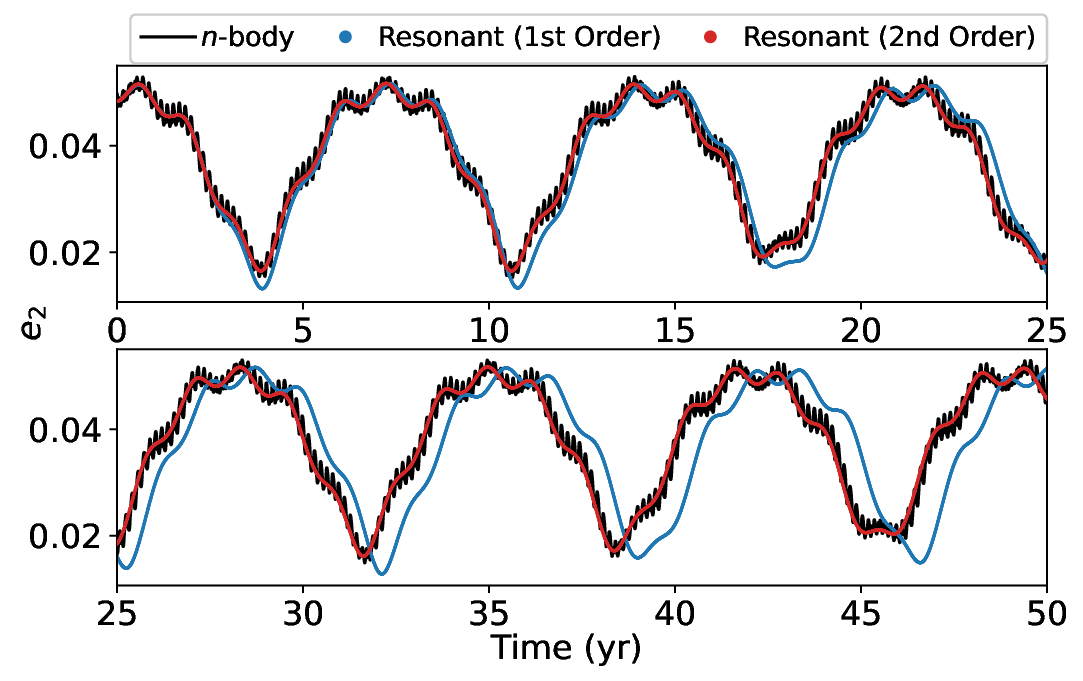}
\caption{Eccentricity evolution of planet c ($e_1$) and planet b ($e_2$) in the GJ 876 system obtained by first-order resonant model (blue), second-order resonant model (red) and direct $n$-body integrations (black).}
\label{fig:GJ876evolution}
\end{figure*}

The eccentricity evolution of planets c and b over 50 years is shown in Fig.~\ref{fig:GJ876evolution}. As in the TIC system, the first-order resonant secular model captures the qualitative behavior of the $n$-body solution but does not accurately reproduce the precession frequencies. Including second-order terms significantly improves the agreement with $n$-body integration, accurately tracking the frequency and amplitude of eccentricity variations. The secular frequencies $g_1$ and $g_2$ are reported in Table~\ref{tab:secular_freq} for each model. At first order, the frequency $g_1$ is reproduced with a relative error of $0.8\%$, but $g_2$ is largely underestimated with a relative error of $112\%$. This highlights the inability of the first-order model to capture the secular dynamics driven by $g_2$. At second order, the error on $g_1$ drops to $0.3\%$, while the error on $g_2$ improves by one order of magnitude, decreasing to $2.6\%$. 
We note that the first order non-resonant secular model does not provide a reliable description of the dynamics in this case, and the second-order formulation in Eq.~\eqref{eq:secularHam_correction} should not be applied either, as the system is locked in a 2:1 MMR.

\section{\label{sec:Conclusion}Discussion}
In this work, we have presented a numerical approach to second-order canonical perturbation theory for the planetary 3-body problem. Our framework is developed using the Lie transform formalism and uses fast Fourier transform to numerically simulate the secular evolution of planetary systems both near and away from MMRs. To validate our approach, we applied it to four distinct planetary systems: the Sun-Jupiter-Saturn (SJS) system, the WASP-148 system, the TIC 279401253 (TIC) system, and the GJ 876 system. 

For the SJS system, our results clearly show that a second-order theory is essential for accurately modeling its long-term evolution. Even with the second-order secular model, we observed a $1 \%$ discrepancy in the precession frequency $g_6$, which we corrected by including the 5:2 MMR terms in the secular Hamiltonian, highlighting the importance of third-order terms in planetary masses associated with this resonance. In the WASP-148 system, despite its proximity to a 4:1 resonance, the second-order non-resonant secular model alone accurately describes the system's evolution. In contrast, the TIC system, being inside a 2:1 MMR, required a resonant model. Incorporating second-order terms here significantly improved the accuracy of the model, though some discrepancies appeared over longer integrations. This is likely to be due to neglecting higher-order terms that can still be important given the high planetary masses in this system, and to the need for a more precise mean-motion correction given the high planet eccentricities (see~Sect.~\ref{par:meanmotion}). In the GJ 876 system, also locked in a 2:1 MMR, the second-order resonant secular model provides a clear improvement over the first-order formulation in capturing the main features of the secular evolution. In particular, it definitively improves the accuracy of the precession frequency $g_2$, reducing the relative error by an order of magnitude compared to the first-order model.

We briefly discuss here the computational cost of our approach, which we implemented in the C programming language. To this end, we compared the integration times and precision of our secular (non-resonant) models with direct $n$-body integration performed using the \texttt{ias15} integrator in \texttt{REBOUND} \citep{Rein2015}. For this comparison, an $n$-body integration of the SJS system over 10 Myr was performed with a fixed time-step of 1/15 years, while the secular models were integrated using an Adams PECE method of order 12 with a fixed time-step of 250 years. We note that, while we selected the high-order \texttt{ias15} scheme to match the accuracy of our Adams integrator, symplectic integrators such as \texttt{WHFAST} or \texttt{SABA1064} are especially suited for long-term $n$-body simulations and can offer substantial speedups depending on the desired precision.
For the secular models, we tested three different grid sizes to perform the FFT: 16 $\times$ 16, 32 $\times$ 32, and 64 $\times$ 64 points. For the second-order secular model, we also tested two finite-difference schemes for the numerical derivatives of the second-order terms in Eq.~\eqref{eq:secularHam_correction} (see Sect.~\ref{par:IntegrationsEOM}): a second-order central difference method and a fourth-order five-point stencil method.
Table~\ref{tab:computational_cost} shows the CPU time, the relative energy error $\Delta \E / \E_0$, and the maximum relative angular momentum error $\Delta \L / \L_0$ obtained for the different integrations. 
The relative errors are defined as:
\begin{equation}
\begin{aligned}
\frac{\Delta \E}{\E_0} &= \frac{|\E(t) - \E_0|}{\E_0}, \\
\frac{\Delta \L}{\L_0} &= \max \left( \frac{|\Delta \L_x|}{\L_0}, \frac{|\Delta \L_y|}{\L_0}, \frac{|\Delta \L_z|}{\L_0} \right),  
\end{aligned}
\end{equation}
where $\E(t)$ is the Hamiltonian evaluated at time t along the integrated solution, and $\E_0$ is the initial energy; the variation over time of the angular momentum components is given by $(\Delta \L_x, \Delta \L_y, \Delta \L_z) = \vec{\L}(t) - \vec{\L}(0)$, where $\vec{\L}(t) = (\L_x(t), \L_y(t), \L_z(t))$ is the total angular momentum vector at time t, and $\L_0$ is the initial angular momentum magnitude. While the energy error primarily reflects the quality of the numerical integration scheme, the angular momentum error provides a more direct measure of the accuracy of the secular models.

For the first-order secular model, using a $16 \times 16$ grid, the orbital solution is computed in a few seconds, almost two orders of magnitude faster than the equivalent $n$-body integration. However, this grid size results in a relatively high angular momentum relative error of $10^{-6}$. Increasing the grid size to $32 \times 32$ and $64 \times 64$ significantly reduces this error to $10^{-10}$ and $10^{-16}$, respectively, at the cost of increased CPU time. Despite this, the first-order model remains faster than the $n$-body integration. It should be noted that the integration time of the first-order secular model can be decisively reduced using Gauss’s method \citep[e.g.,][and references therein]{Mogavero2021}. 
 
At order two, the computational cost of the secular model increases due to the computation of the partial derivatives of $\widehat{H}_2$ required in the equations of motion. On top of this, using a numerical derivative scheme adds an additional cost, as it involves several evaluations of the function $\widehat{H}_2$. We have reduced the cost of differentiation by parallelizing the calculation of the partial derivatives of $\widehat{H}_2$. The resulting second-order secular model is either slightly faster or slower than the $n$-body model, depending mainly on the FFT grid size. As shown in Table~\ref{tab:computational_cost}, the five-point stencil method achieves greater precision compared to the central difference method, especially when using finer FFT grid sizes. For example, with a $64 \times 64$ grid, the maximum relative angular momentum error decreases from $10^{-11}$ (central difference) to $10^{-14}$ (five-point stencil). In contrast, for coarser grids, the choice of the numerical derivative scheme has a minimal impact, with errors remaining at $10^{-6}$ and $10^{-10}$ for $16 \times 16$ and $32 \times 32$ grid sizes respectively. Energy conservation is largely unaffected by grid size for both the first- and second-order models, though it is slightly influenced by the choice of the numerical derivative scheme at second order.
\begin{table*}
\centering
\caption{\textbf{Computational cost and precision for a 10-Myr integration of the SJS system.} Relative energy error ($\Delta \E / \E_0$) and maximum relative error in the angular momentum components ($\Delta \L / \L_0$) are shown for different models alongside CPU time and FFT grid size. For the second-order secular model, we test two finite difference schemes for numerical derivatives: central difference (second order) and five-point stencil method (fourth order). All calculations were performed with an Apple M1 Pro chip (10-core CPU).}
\label{tab:computational_cost}
\resizebox{0.6\textwidth}{!}{%
\begin{tabular}{r|c|S[table-format=2.1]|c|c}
\hline \hline 
Model & Grid Size & {CPU Time (min)} & $\Delta \E/\E_0$ & $\Delta \L/\L_0$ \\ \hline 
$n$-body &  & 8.0 & $10^{-14}$ & $10^{-13}$\\ \hline
Secular 1st & $16 \times 16$ & 0.1 & $10^{-16}$ & $10^{-6}$ \\
                  & $32 \times 32$ & 0.4 & $10^{-16}$ & $10^{-10}$ \\
                  & $64 \times 64$ & 1.8 & $10^{-16}$ & $10^{-16}$ \\ \hline
Secular 2nd & $16 \times 16$ & 0.6 & $10^{-14}$ & $10^{-6}$ \\
(Central difference)& $32 \times 32$ & 2.0 & $10^{-14}$ & $10^{-10}$ \\
                   & $64 \times 64$ & 9.3& $10^{-14}$ & $10^{-11}$ \\\hline
Secular 2nd &  $16 \times 16$  & 1.0 & $10^{-16}$ & $10^{-6}$\\
(Five-point stencil)& $32 \times 32$  & 3.6 & $10^{-16}$ & $10^{-10}$ \\
                    & $64 \times 64$  & 14.6& $10^{-16}$ & $10^{-14}$ \\ \hline
\end{tabular}}
\end{table*}

The important factor that affects the computational cost of our approach is the number of grid points used in the FFT, as evaluating the disturbing function and its first-order analytical derivatives is the most expensive step in our implementation. As shown in Table~\ref{tab:computational_cost}, reducing the FFT grid from 64 $\times$ 64 to 32 $\times$ 32 points decreases the computational cost by a factor of 4, although it increases the relative angular momentum error to $10^{-10}$ compared to $10^{-16}$ and $10^{-14}$ for the first-order model and the second-order model, respectively, with the five-point stencil method. This trade-off between precision and computational cost is thus an important consideration when selecting the grid size for a given system. Another crucial factor affecting the speed of secular integrations is clearly the timescale separation between the orbital periods and the secular precession frequencies, and this strongly depends on the planetary system considered.

Overall, our results show that the inclusion of second-order terms in secular models significantly improves the accuracy of orbital precession frequencies, particularly in systems affected by strong MMRs or where the planetary masses are large. These frequencies are crucial because they possibly determine the chaotic character of the secular dynamics, as is the case in the solar system. Our numerical approach, free from limitations typically associated with truncated expansions in eccentricities, inclinations, or semi-major axis ratios, enables the exploration of a wide variety of exoplanetary systems, allowing us to isolate the main planet interactions that drive their secular dynamics, and offering deeper insights into their long-term stability.

\begin{acknowledgments}
We are grateful to the anonymous reviewer for a detailed report and valuable comments that helped to improve the manuscript. 
F.M. has been supported by a grant of the French Agence Nationale de la Recherche (AstroMeso ANR-19-CE31-0002-01). This project has been supported by the European Research Council (ERC) under the European Union’s Horizon 2020 research and innovation program (Advanced Grant AstroGeo-885250). This work was granted access to the HPC resources of MesoPSL financed by the Region Île-de-France and the project Equip@Meso (reference ANR-10-EQPX-29-01) of the  programme Investissements d’Avenir supervised by the Agence Nationale pour la Recherche. 
\end{acknowledgments}


\bibliography{biblio}

\end{document}